\documentclass[12pt,preprint]{aastex}

\shorttitle{The Asiago-ESO/RASS QSO Survey II}
\shortauthors{Grazian et al.}

\begin{document}


\title{The Asiago-ESO/RASS QSO Survey \\
II. The Southern Sample\footnote{Based on observations collected at the
European Southern Observatory, Chile (ESO P66.A-0277 and ESO P67.A-0537),
with the Arizona Steward Observatory
and with National Telescope Galileo (TNG) during AO3 period.}}

\author{A. Grazian}
\affil{European Southern Observatory, D-85748 Garching bei M\"{u}nchen
\and
Astronomy Department, University of Padua, I-35122 Italy}
\email{agrazian@eso.org, grazian@pd.astro.it}

\author{A. Omizzolo \and C. Corbally}
\affil{Vatican Observatory Research Group, University of Arizona,
             Tucson AZ 85721 USA}
\email{aomizzolo@specola.va, ccorbally@as.arizona.edu}

\author{S. Cristiani}
\affil{European Southern Observatory, ST European Coordinating
             Facility, D-85748 Garching bei M\"{u}nchen
\and
Osservatorio Astronomico di Trieste
             Via G.B. Tiepolo 11, 34131 Trieste Italy}
\email{scristia@eso.org, cristiani@ts.astro.it}

\author{M. G. Haehnelt}
\affil{Institute of Astronomy, Madingley Road, Cambridge CB30HA, England}
\email{m.haehnelt@ic.ac.uk}

\and

\author{E. Vanzella}
\affil{European Southern Observatory, D-85748 Garching bei M\"{u}nchen
\and
Astronomy Department, University of Padua, I-35122 Italy}
\email{evanzell@eso.org, vanzella@pd.astro.it}



\begin{abstract}
This is the second paper of a series describing the Asiago-ESO/RASS QSO
survey, a project aimed at the construction of an all-sky
statistically well-defined sample of very bright QSOs ($B_J\le
15$). Such a survey is required to remove the present uncertainties about the
properties of the local QSO population and constitutes an homogeneous
database for detailed evolutionary studies of AGN.
We present here the complete Southern Sample, which comprises 243 bright
($12.60\le B_J\le 15.13$) QSO candidates at high galactic latitudes
($|b_{gal}|\ge 30^{\circ}$). The area covered by the survey is 5660 sq. deg.
Spectroscopy for the 137 still unidentified objects has been obtained.
The total number of AGN turns out to be 111, 63 of which are new
identifications. The properties of the selection are discussed.
The completeness and the success rate for this survey at the final stage
are $63\%$ and $46\%$, respectively.

\end{abstract}




\section{Introduction}

QSOs are an important astrophysical and cosmological tool: 
they represent a major source of information about the origin and evolution of
the structures in the Universe. They can be used either directly, as tracers
of the density peaks, or as cosmic lighthouses, probing the Universe along the
line of sight with their conspicuous flow of photons.

Early stages of galaxy
formation are probably connected with QSO activity and central BH accretion.
In recent years there has been increasing observational evidence that
the evolution of normal galaxies and quasars is closely related and
that quasars are short-lived. The evolution of the global star
formation rate of the Universe, the space density of
starbursting galaxies and that of luminous QSOs appear to be
remarkably similar. Recently \citet{korm1a} proposed that such
``monsters'' could be set at the heart of galaxy formation.
A number of models indeed relate QSOs with galaxies both using
theoretical \citep{HK2000,granato,monaco,romano} and observational
\citep{gebha,korm01} arguments. It is generally accepted now that
QSO activity, the growth of SMBHs and the formation of spheroids are
all closely linked \citep{korm1b}.

A well defined large sample of bright QSOs at $z\le 0.3$
is instrumental in confirming or revising our conceptions about the
evolution of QSOs and constitutes a significant challenge for any theoretical
model. In particular, it provides key informations on the following issues:
{\em what is the typical mass of Dark Matter Halos hosting AGN ?
What is the duty cycle for AGN activity ? What is the typical efficiency of
the central engine at the various redshifts ?}

It is paradoxical that in the era of 2dF \citep{croom01,2df}
and SDSS \citep{sdss}, with
thousands of faint QSOs discovered up to the highest redshifts ($z=6.28$),
the statistical properties of the QSO population are much better known at
$z\sim 2$ than in the local Universe.
The aim of this work is to fill this gap with a very large area search for
bright and low-z AGN, the Asiago-ESO/RASS QSO Survey (hereafter AERQS).
At present, the Northern Sample described in \citet{grazian}
(hereafter Paper I) is 85\%
identified and spectroscopic observations have been planned to complete
the survey.

The structure of this paper is the following: in \S\ 2 we report on the
Southern Photometry used, the selection criteria are
described in \S\ 3; \S\ 4 is dedicated to
the New Southern Sample and its statistical properties (selections,
completeness, efficiency); finally on \S\ 5 we discuss the properties of the
completely identified sample, showing the spectra of newly identified
AGN and galaxies.
We assume $H_{0}=50~Km~s^{-1}~Mpc^{-1}$, $\Omega_{m}=1.0$ and
$\Omega_{\Lambda}=0.0$.
A detailed treatment of the LF and clustering properties of AGN is left to
forthcoming papers.

\section{The Southern Photometry}

The surface density of bright AGN at low redshift is very small, around 
$10^{-2} sq.~deg.^{-1}$ as shown in Fig. \ref{lognlogs}. In practice,
to reach significant statistics, an area comparable with the whole sky
has to be covered.
In Paper I we have discussed a number of databases
sampling a wide domain in the electro-magnetic (e.m.) spectrum for the
selection of an all-sky sample of optically bright QSOs with a high level of
completeness and success rate.
\placefigure{lognlogs}

In the northern part of the AERQS the basis of the optical photometry
was chosen to be the US Naval Observatory Catalogue
(USNO-1A\footnote{\sl \url{http://archive.eso.org/servers/usnoa-server}})
and the Guide Star Catalogue
(GSC-1\footnote{\sl \url{http://www-gsss.stsci.edu/gsc/gsc12/}})
with a typical error of 0.2-0.3 in magnitude. For the Southern Sample we
have tried to improve the optical photometry, using
the positions and optical magnitudes derived from the Digitized
Sky Survey (DSS\footnote{\sl \url{http://archive.eso.org/dss/dss}}).
For each target of interest (the selection is described in \S\ 3)
all the objects with known $B_J$ magnitudes within a radius of 1.5 arcmin
were extracted from the GSC Catalogue.
Small scans of the target and the GSC calibrating objects were extracted
from the DSS plates.
Instrumental magnitudes were then computed
by aperture photometry in a circular area of 7.5 arcsec radius (9 DSS pixels
diameter). A polynomial calibration curve is used to
interpolate the magnitudes of the target.
A typical calibration curve is shown in Fig. \ref{calib}.
We have tested the accuracy of this procedure using 446 photometric
standards of the input catalogue used to calibrate the photometric
material of the Homogeneous Bright QSO Survey \citep{hbqs}, deriving
a $\sigma_{B_J}$ of 0.10 $mag$ in the interval $12.0\le B_J\le 15.5$.
\placefigure{calib}

The 7.5 arcsec aperture size (corresponding to a radius of 18.5 Kpc
at $z=0.1$) is the result of a trade-off between
the attempt to estimate nuclear
magnitudes for our AGN (reducing the contribution of the host galaxy) and the
necessity of a photometry that is ``robust'' against errors in the centering
of the aperture. By comparing our photometry with the GSC2 catalogue we find,
for the 80 AGN of Tab. \ref{dsssample} that have GSC2 $J$ magnitudes, a mean
difference $<J_{GSC2}-B_J>=0.10$ with a scatter of 0.4 mag, which
can be ascribed to photometric errors and AGN variability.
Using larger apertures obviously increases the contribution of the host
galaxy. For example
if we compare the magnitudes obtained with a 7.5 arcsec circular aperture
with the magnitudes in a 15 arcsec aperture, for the 111 AGN
of Tab. \ref{dsssample} we obtain a $<B_J(7.5)-B_J(15.0)>=0.5$ with a
scatter of 0.4.

We have used only plates based on {\em IIIaJ} emulsion to compute the
magnitudes, as other emulsions are not standard and difficult to calibrate.
This, together with the selection criteria described below, is
the source of non uniformity of the sky coverage.
Fig. \ref{areadss} shows the area of the sky covered by the present survey.
Table \ref{areasud} provides a list of the sky sub-areas plotted in
Fig. \ref{areadss}, which total 5660 sq. deg. of the Southern Hemisphere.
\placefigure{areadss}
\placetable{areasud}

\section{The Selection Criteria}

QSO candidates have been selected by cross correlating the X-ray sources in the
ROSAT All Sky Survey Bright Source Catalog (RASS-BSC, \citet{vog99})
with optically bright objects in the DSS plates.
As stated in Paper I, given the low surface density of local bright AGN,
misidentifications are very unlikely since we adopt a matching radius
that is three times the RMS positional uncertainty of each X-ray source
(typically 15-20 $arcsec$).

We want to stress here that this survey aims at finding
{\em optically bright} QSOs and the X-ray emission
is used only to compute an ``X-Optical color'' for the selection of AGN.
Therefore the result of our selection cannot be
considered an identification of X-ray sources.
This makes the follow-up spectroscopy
quicker than in the case of optical identifications of X-ray sources,
because we do not care about objects fainter than the chosen optical flux
limits and mis-identifications of optically fainter X-ray sources have no
effect on the result.

We applied to the X-ray catalogue a number of criteria that do not
affect drastically the completeness of our survey (basically the same used in
Paper I): exposure time $t_{exp}\ge 300 s$, Galactic latitude
$|b_{gal}|\ge 30^{\circ}$, hardness ratio in the 0.5$\div$2.0 and
0.1$\div$0.4 keV energy bands $-0.9\le HR1\le +0.9$, hardness ratio in the
0.9$\div$2.0 and 0.5$\div$0.9 keV energy bands $-0.6\le HR2\le +0.8$,
likelihood of extent
$lik_{ext}\le 35$ to avoid extended X-ray sources; this corresponds
to a limit for source extent $ext\le 100$ in agreement with the preliminary
results of the North Ecliptic Pole (NEP) survey
(\citet{vog01} and C. Mullis private communications, 2001).
In addition the likelihood of detection $lik_{det}\ge 25$ to select only
reliable sources, with a significant level of detection in the RASS-BSC.
These parameters have been described extensively in \citet{vog99}.

Then we apply two basic criteria:

$\bullet$ $12.60\le B_J\le 15.13$

and

$\bullet$ $a_{ox}\le 1.9$

where
$a_{ox}=-0.438log_{10}(cps)-0.193B_J+4.20$ and $cps$ is the X-ray flux
measured in counts per second.

We have used the selection criterion 
$\alpha_{ox}\le \alpha_{max}$ which, as shown in Fig. 1 of Paper I,
for objects brighter
than the adopted optical limits provides a sample with a degree of
incompleteness that is not a function of the apparent magnitude.
We have compared the present selection with the low redshift ($z\le 0.3$)
optically  or IR selected QSOs of the V\'eron Catalogue (\citet{VV01},
hereafter VV01): out of the 67 QSOs
known within our spatial and optical flux limits, 42 (63 $\%$)
meet our selection criteria.
Radio or X-ray selected AGN are not taken into account, to avoid
biases in the estimation of the completeness.

The adopted selections in $lik_{det}$, $lik_{ext}$, $HR1$ and $HR2$
remove $25\%$ of the RASS sources that are probably stars, extended X-ray
sources and other spurious contaminants; spectroscopic identifications for
these sources are not available. The application of the same criteria for the
AGN in the VV01 Catalogue lowers the completeness from $64\%$ to $63\%$;
we can conclude, as in Paper I, that the adopted criteria increase the
effectiveness without affecting the completeness.

We have selected a total of 243 candidates
in the Southern Hemisphere over $\sim$5660 $sq.~deg.$ at high Galactic latitude
$|b_{gal}|\ge 30^{\circ}$. They are listed in Tab. \ref{dsssample}.
\placetable{dsssample}

\section{The Southern Sample}

Of the 243 candidates belonging to the southern part of the AERQS, 45\%
had previous spectroscopic identifications in the literature (V\'eron
Catalogue, NED\footnote{\sl \url{http://nedwww.ipac.caltech.edu/}}).
For the remaining 137 objects we started an observational campaign. We
had several runs with different telescopes for a total of 7 nights:
Tab. \ref{jobs} summarizes the observations.
\placetable{jobs}

The reduction process used the standard MIDAS facilities \citep{midas}
and other useful software available at ESO Garching through
the SCISOFT\footnote{\sl \url{http://www.eso.org/science/scisoft/}}
environment.
The raw data were sky-subtracted and corrected for pixel-to-pixel sensitivity
variations by division with a suitably normalized exposure of the spectrum of
an incandescent source (flat-field).
The wavelength calibration was carried out by
comparison with exposures of He and Ne lamps. Relative flux
calibration was carried out by observations of spectrophotometric
standard stars \citep{oke}.
For extended objects, only the core/nucleus flux was considered.

The identification classes reported in Table \ref{dsssample} are:
$AGN$ = Active Galactic Nucleus;
$STAR$ = star; $GAL$ = galaxy; $BLLAC$ = BL Lac object.

Objects with emission lines were classified as AGN only if they show
broad and/or strong lines (typically $H_{\alpha}$, $H_{\beta}$ or
\ion{Mg}{2}).
Galaxies with a weak ($EW\le 12\AA$) unresolved $H_{\alpha}$
and no other features of AGN activity are classified
as $EM~GAL$ and, together with the newly identified AGN or normal galaxies,
are shown in
Fig.\ref{spec1},\ref{spec2},\ref{spec3},\ref{spec4},\ref{spec5},\ref{spec6}.
The objects classified as BL Lacs in Tab. \ref{dsssample} were already
known from the literature (VV01 Catalogue).
In the next section we will describe more in detail the emission line
galaxies and try to interpret their properties.
\placefigure{spec1}
\placefigure{spec2}
\placefigure{spec3}
\placefigure{spec4}
\placefigure{spec5}
\placefigure{spec6}

At the end of our spectroscopic campaign we have carried out
137 new identifications; we have discovered more than 60 new AGN,
significantly
enlarging the number of bright QSOs at $z\le 0.3$. Fig. \ref{histz} shows
the redshift distribution of the AGN in this sample.
\placefigure{histz}

\section{Discussion}

We have found 111 AGN out of 243 candidates, corresponding to a success rate
of 46$\%$.
Stars are the mean source of contamination, especially active M stars,
which are
powerful X-ray emitters compared to their optical magnitudes, resembling
the $a_{ox}$ of AGN. To distinguish them effectively an optical color, for
example $B-R$, would be extremely useful as these two classes have
typically different optical spectral energy distributions.
We have obtained $J$ and $F$ magnitudes,
equivalent to $B$ and $R$ respectively,
from the GSC-2\footnote{\sl
\url{http://www-gsss.stsci.edu/gsc/gsc2/GSC2home.htm}} catalogue
for all the object of this survey. In Fig. \ref{histbr} the $J-F$ color
distribution is plotted for different classes of objects.
We have divided AGN into two classes ``Point-like'' and ``Extended''
or ``Galaxy-like'' according to the classification given in the GSC-2
catalogue.
There is no evident difference in colors between these two classes.
AGN and normal stars are not so different in $J-F$.
M-stars, instead, can be easily separated from AGN.
With the application of a reasonable cut in the optical color ($J-F\le 1.6$)
the success rate of the present survey would be increased from the present
value of 46$\%$ to 63$\%$ but the completeness would be affected as well,
decreasing from 63$\%$ to a value of 40$\%$.
If we compare the ``Extended'' and ``Point-like'' AGN of
Fig. \ref{histbr}, a Kolmogorov-Smirnov test gives a probability of $89\%$
that the two samples are extracted from the same population. The mean values
of $J-F$ for the two samples are similar (0.69 and 0.72 for ``Extended'' and
``Point-like'', respectively) and the dispersion is slightly
larger for the ``Extended'' objects.
\placefigure{histbr}

A more important
consideration is the fact that surveys based only on optical colors, assuming
typical blue SEDs for AGN, are significantly incomplete especially at low
redshift and at faint absolute magnitudes, where the host galaxy contribution
starts to be relevant. Fig. \ref{mabsjf} shows the dependence of the
AGN color $J-F$ on absolute magnitude $M_{B}$: faint Nuclei tend to be redder
than the bright QSOs.
In Fig. \ref{histpg} we show the $J-F$ color distribution for 30 QSOs with
$z\le 0.3$ in the PG Survey \citep{PG83}.
We compare it with the same
distribution for 80 AGN in the AERQS Survey with $z\le 0.3$: an extended
tail towards the red $J-F$ color for the X-ray selected AGN is evident.
PG QSOs have typically a blue optical color
($J-F\le 1.04$). If we had selected only AGN bluer than $J-F\le 1.04$, 
22 ($28\%$) objects would have been missed.

Two effects can determine the big spread in the observed $J-F$ color:
the starlight contamination of the host galaxy and the existence of
intrinsically red Active Galactic Nuclei.
An additional contribution can be due to QSO variability, whose effect is
difficult to address in detail, as it significantly depends on the time lag
between the different flux measurements.
From the analysis of the structure function \citep{variability} we should
expect an average uncertainty on the $J-F$ color due to variability of 0.2
mag for QSOs with a typical absolute magnitude $M_B\sim -25$ and 0.3 for
$M_B\sim -23$.

The contribution of the host galaxy is clearly visible in
Fig. \ref{composite}, where we have normalized and stacked all QSO spectra
obtained in this survey. We compare the result with the composite spectra
by SDSS \citep{VBSDSS}, First Bright Quasar Survey \citep{FBQS} and with
a synthetic spectrum used for photometric
redshift studies with a continuum slope of $\nu^{-1.75}$, redder than a
typical $\nu^{-1.2}$ Blue QSOs. It is clearly visible in our composite
spectrum the red continuum and the strong feature typical of early type
galaxies (Ca doublet at 3929.3 and 3963.8 \AA), producing a significant
absorption in the rest-frame B band.
Besides, it is apparent that for QSOs fainter than $M_B=-24$ the contribution
of the host galaxy produces a redder SED with respect to QSOs brighter than
$M_B=-24$.
K-corrections are computed following the recipe of \citet{kcorr90},
but based on the new QSO composite spectra (FBQS and SYNT) plotted in
Fig. \ref{composite}.

In Fig. \ref{fbqsredell} we have tried to model the pattern of $J-F$ color
observed in Fig. \ref{mabsjf}.
To reproduce the full range in $J-F$ color, both a contamination
from the host galaxy and the existence of AGN bluer and redder than the
adopted composite spectra are necessary. QSO variability and photometric
errors are expected to increase the scatter observed in Fig. \ref{mabsjf}
with respect to Fig. \ref{fbqsredell}. Clearly the synthetic QSO spectrum
is too red with respect to the observed $J-F$ distribution, while
the FBQS composite spectrum is roughly in agreement with the observations
(a slightly bluer QSO spectrum would produce an even better match).
A morphological analysis of individual cases is required in order to
quantify the relative incidence of these effects.
\placefigure{mabsjf}
\placefigure{histpg}
\placefigure{composite}
\placefigure{fbqsredell}

There are 5 objects with $H_{\alpha}$ in emission, faint [\ion{O}{3}]
doublet and no other signature of AGN activity.
We have
classified them as $EM~GAL$ in Table \ref{dsssample}. They could be
special cases, for example AGN obscured by dusty torus, according to the
unified model.
Another possibility is that they are normal starbursts or
liners, common in a soft X-ray survey like the ROSAT sample. Further analysis,
for example using hard X-ray observations with Chandra or XMM-Newton,
can shed light on
their nature and disentangle between Starburst and AGN activity.
In the following papers only objects classified as bona-fide $AGN$ will be
taken into account to study properties like clustering or Luminosity Function.

\begin{figure}
\epsscale{0.80}
\plotone{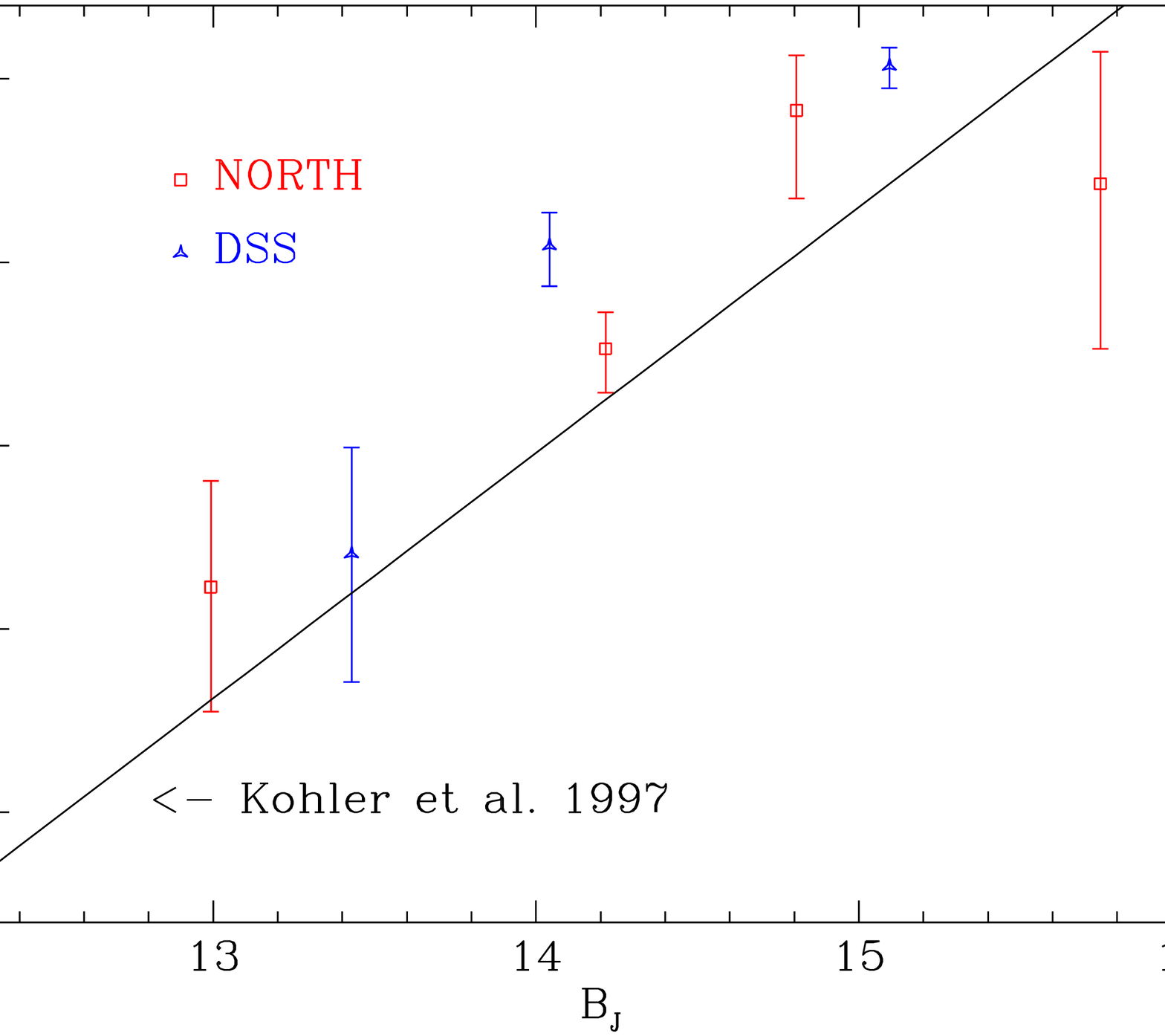}
\caption{The LogN-LogS relation of QSOs. Triangles represent the
current sample and are AGN with $0.04\le z\le 0.2$. The open squares are
the analogs for the northern part of the AERQS \citep{grazian}.
The solid line is the
relation found by K\"{o}hler et al. (1997) for QSOs with $0.7\le z\le 2.2$.
\label{lognlogs}}
\end{figure}

\begin{figure}
\epsscale{0.80}
\plotone{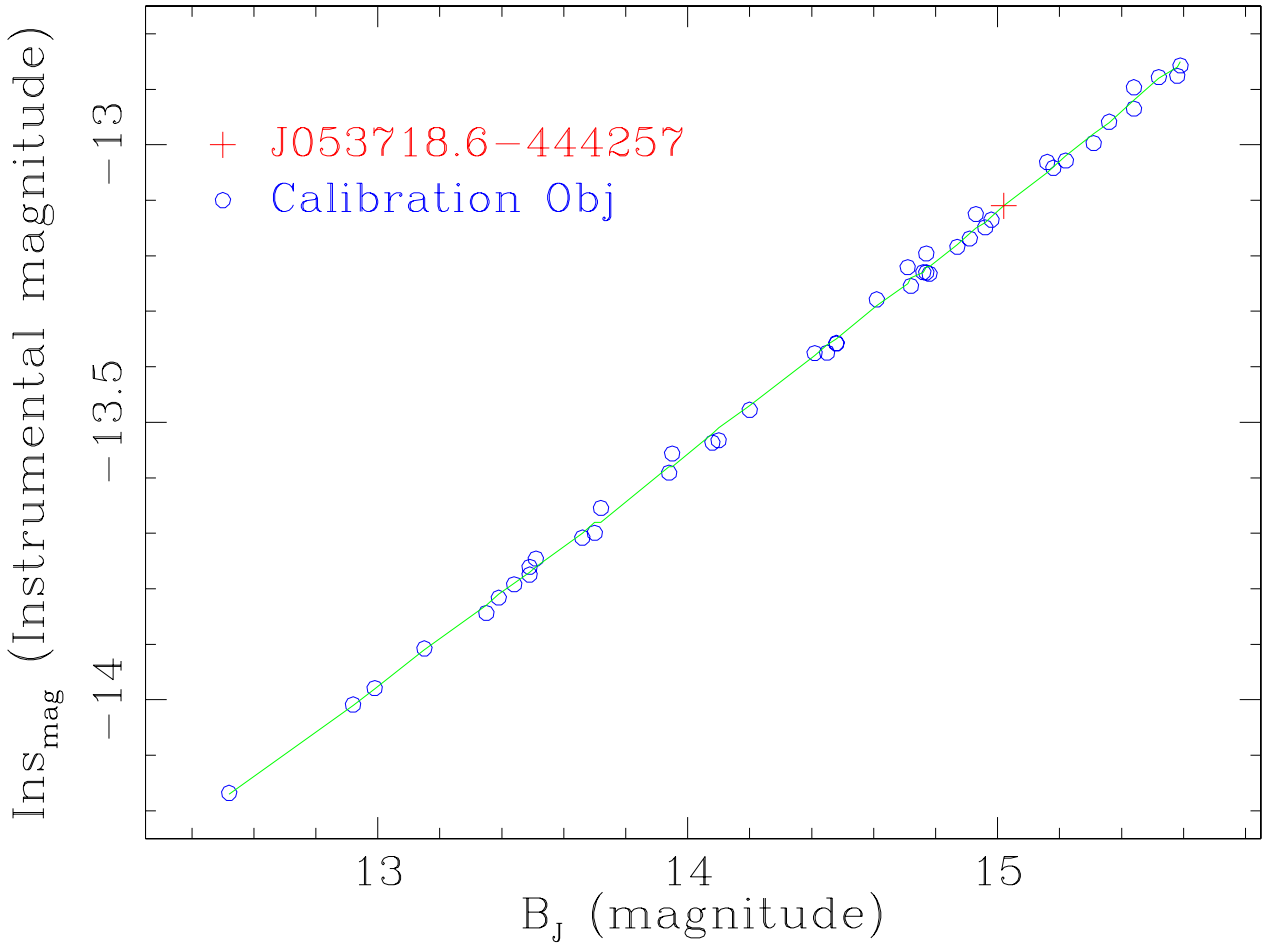}
\caption{A typical calibration curve for one of the AERQS candidates
(1RXS-J053718.6-444257). 45 objects with known $B_J$ magnitude from
the GSC Catalogue within a radius of 1.5 arcmin. were used to derive
the calibration curve.
\label{calib}}
\end{figure}

\begin{figure}
\plotone{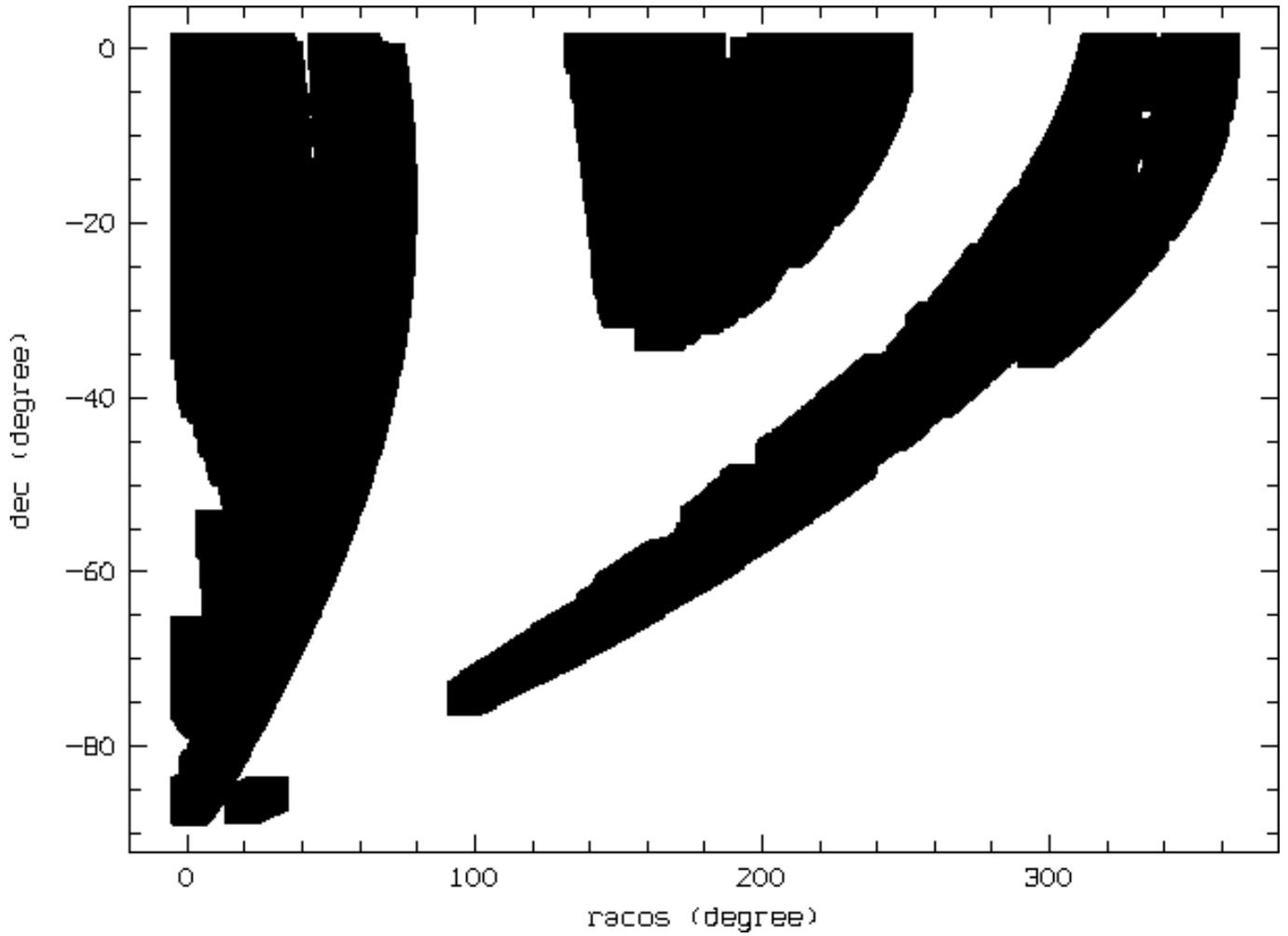}
\caption{The black area shows the regions covered by the present sample
after taking into account the selection criteria described in section 2 and 3.
The projection of the southern sky is done here in $RA\cos(DEC)$ vs
$DEC$.
\label{areadss}}
\end{figure}

The LogN-LogS relation for AGN belonging to this sample is shown in
Fig. \ref{lognlogs} and compared with the relation found by \citet{kohler}
for QSOs with $0.07\le z\le 2.2$. It is also consistent with the same
relation found for the northern part of the AERQS.

With the completion of the southern part of the AERQS a statistically
well-defined set of 340 bright QSOs with $z\le 0.3$ has been collected.
{\bf On the basis of the measured success rate, at the end of the present
project, we expect to provide a full-sky ``local'' sample of 400 AGN.}

\acknowledgments
We warmly thank the referee for carefully reading the manuscript,
for useful suggestions and for improving significantly the quality of this
paper.
Part of the work has been supported by the European Community 
Research and Training Network "Physics of the Intergalactic Medium".
AG was supported by the ESO DGDF 2000 and by an ESO Studentship and
acknowledges the generous hospitality of ESO headquarters during his stay
at Garching.
It is a pleasure to thank R. Mignani for his invaluable help with the GSC-2
and A. Goncalves Darbon for her precious suggestions on objects classifications
and interesting discussions.
This project has been supported by the European Commission through the
``Access to Research Infrastructures Action of the Improving Human
Potential Programme'', awarded to the 'Instituto de Astrof\'isica de
Canarias' to fund European Astronomers access to the European Northern
Observatory, in the Canary Islands.
This paper makes use of the ROSAT All-Sky Survey Bright Source Catalogue
(1RXS).
The Guide Star Catalogue-II (GSC-2) is a joint project of the Space Telescope
Science Institute and the Osservatorio Astronomico di Torino.
Space Telescope Science Institute is operated by the Association of
Universities for Research in Astronomy, for the National
Aeronautics and Space Administration under contract NAS5-26555.
The participation of the Osservatorio Astronomico di Torino is
supported by the Italian Council for Research in Astronomy. Additional
support is provided by European Southern Observatory, Space
Telescope European Coordinating Facility, the International GEMINI project
and the European Space Agency Astrophysics Division.
Based on photographic data obtained using the UK Schmidt Telescope.
The UK Schmidt Telescope was operated by the Royal Observatory Edinburgh,
with funding from the UK Science and Engineering Research
Council, until 1988 June, and thereafter by the Anglo-Australian Observatory.
Original plate material is copyright the Royal Observatory Edinburgh and
the Anglo-Australian Observatory. The plates were
processed into the present compressed digital form with their permission.
The Digitized Sky Survey was produced at the Space Telescope Science
Institute under US Government grant NAG W-2166.
This research has made use of the NASA/IPAC Extragalactic Database (NED)
which is operated by the Jet Propulsion Laboratory, California Institute of
Technology, under contract with the National Aeronautics and Space
Administration.

\clearpage


\begin{figure}
\plotone{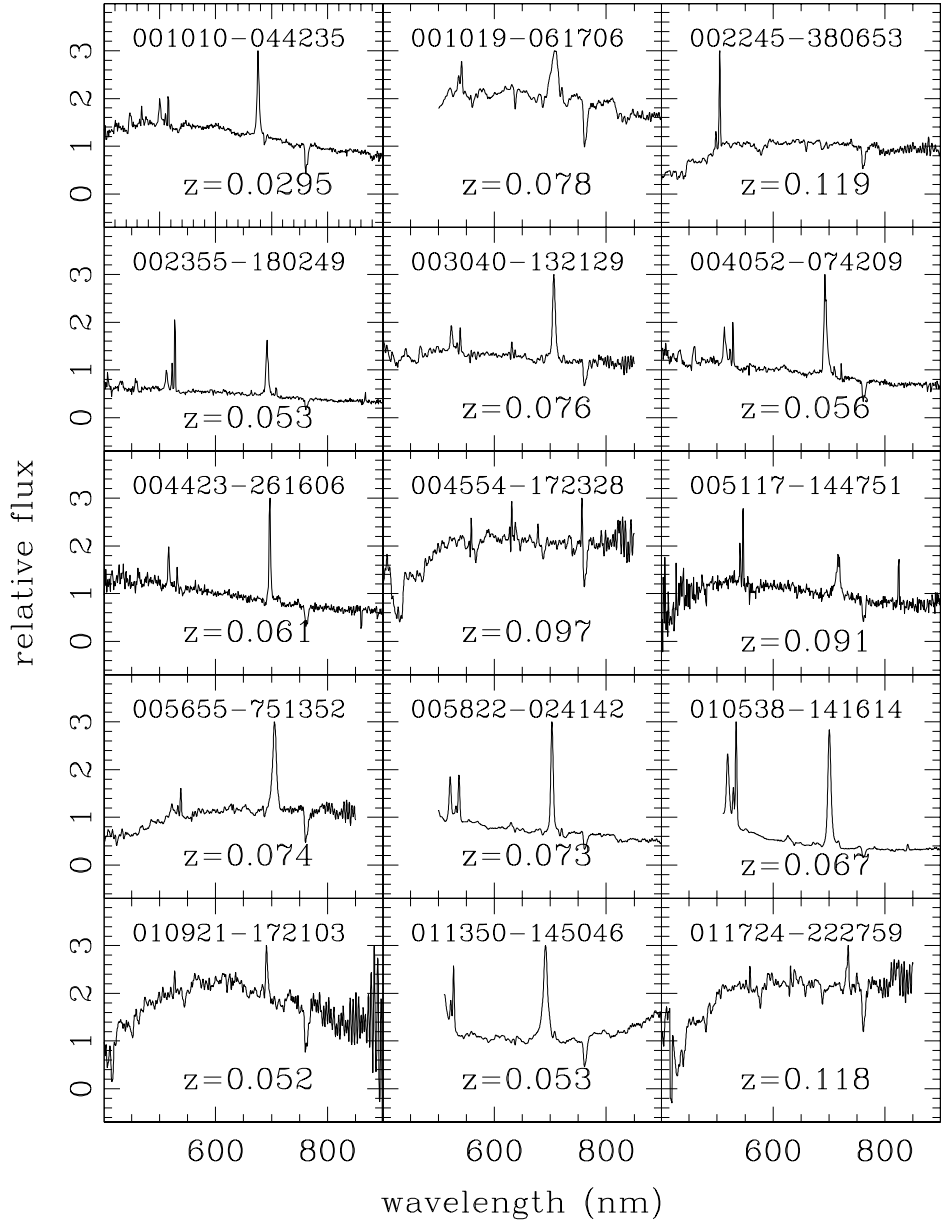}
\caption{Spectra of the newly identified AGN and Galaxies during the present
survey.\label{spec1}}
\end{figure}

\begin{figure}
\plotone{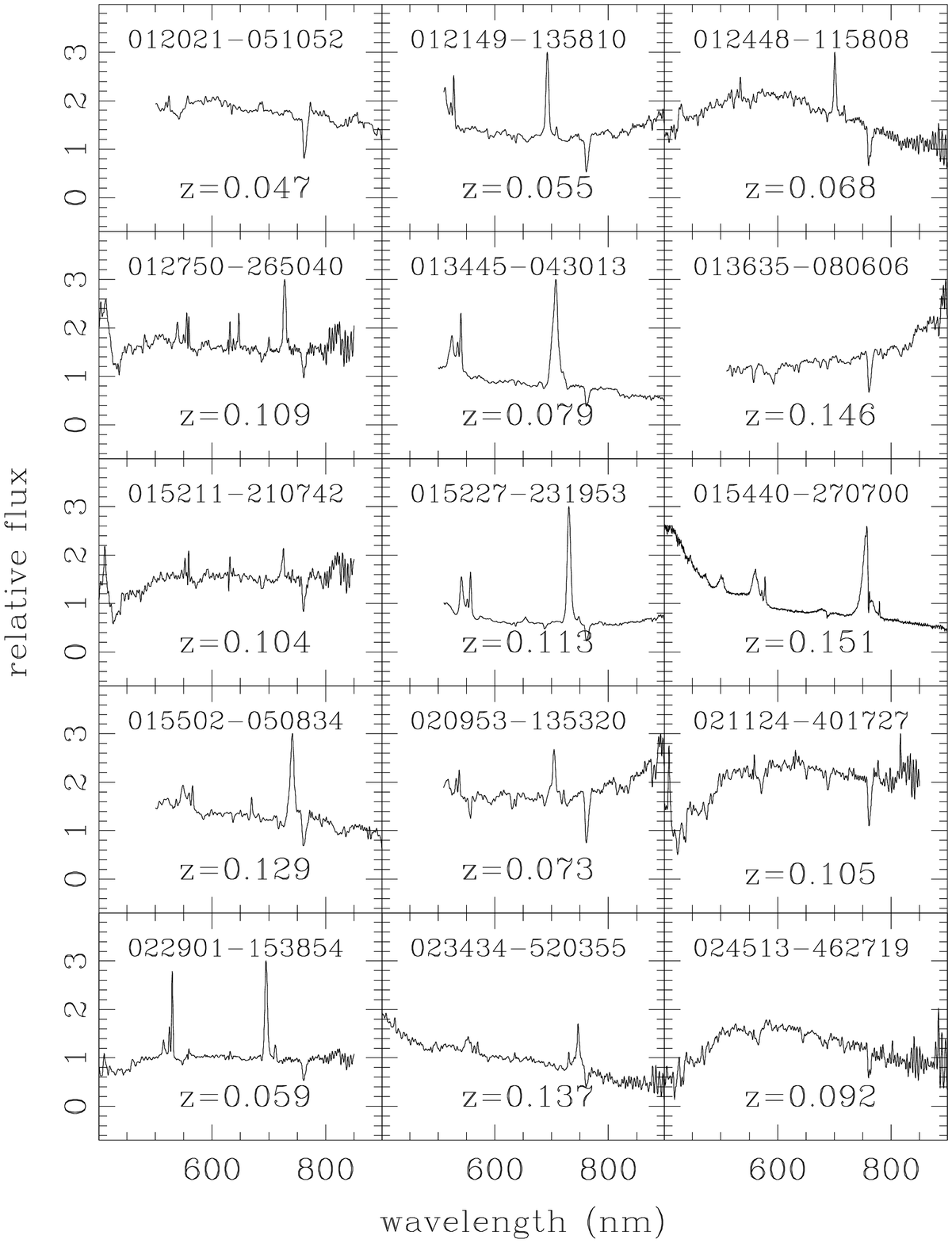}
\caption{Continued.\label{spec2}}
\end{figure}

\begin{figure}
\plotone{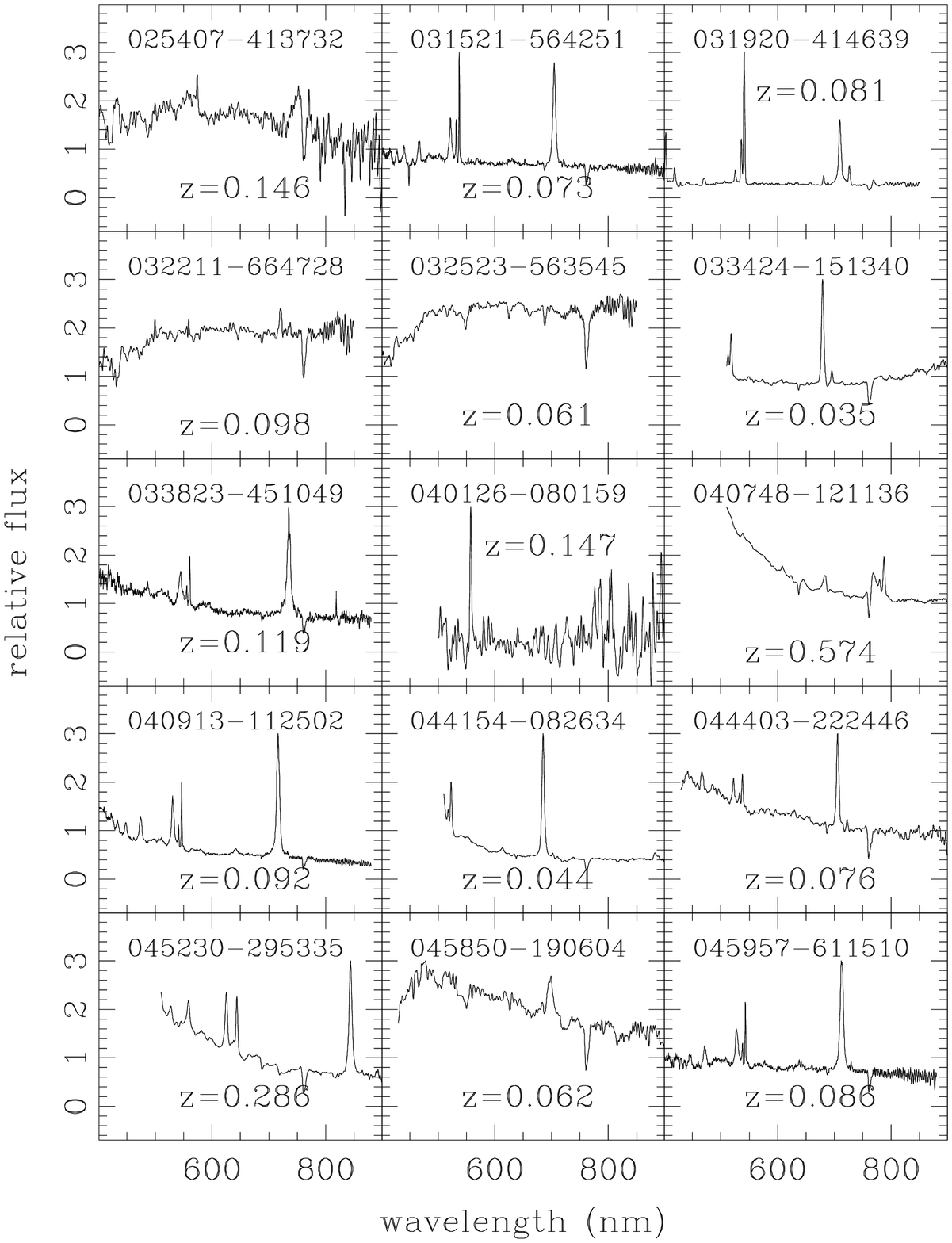}
\caption{Continued.\label{spec3}}
\end{figure}

\begin{figure}
\plotone{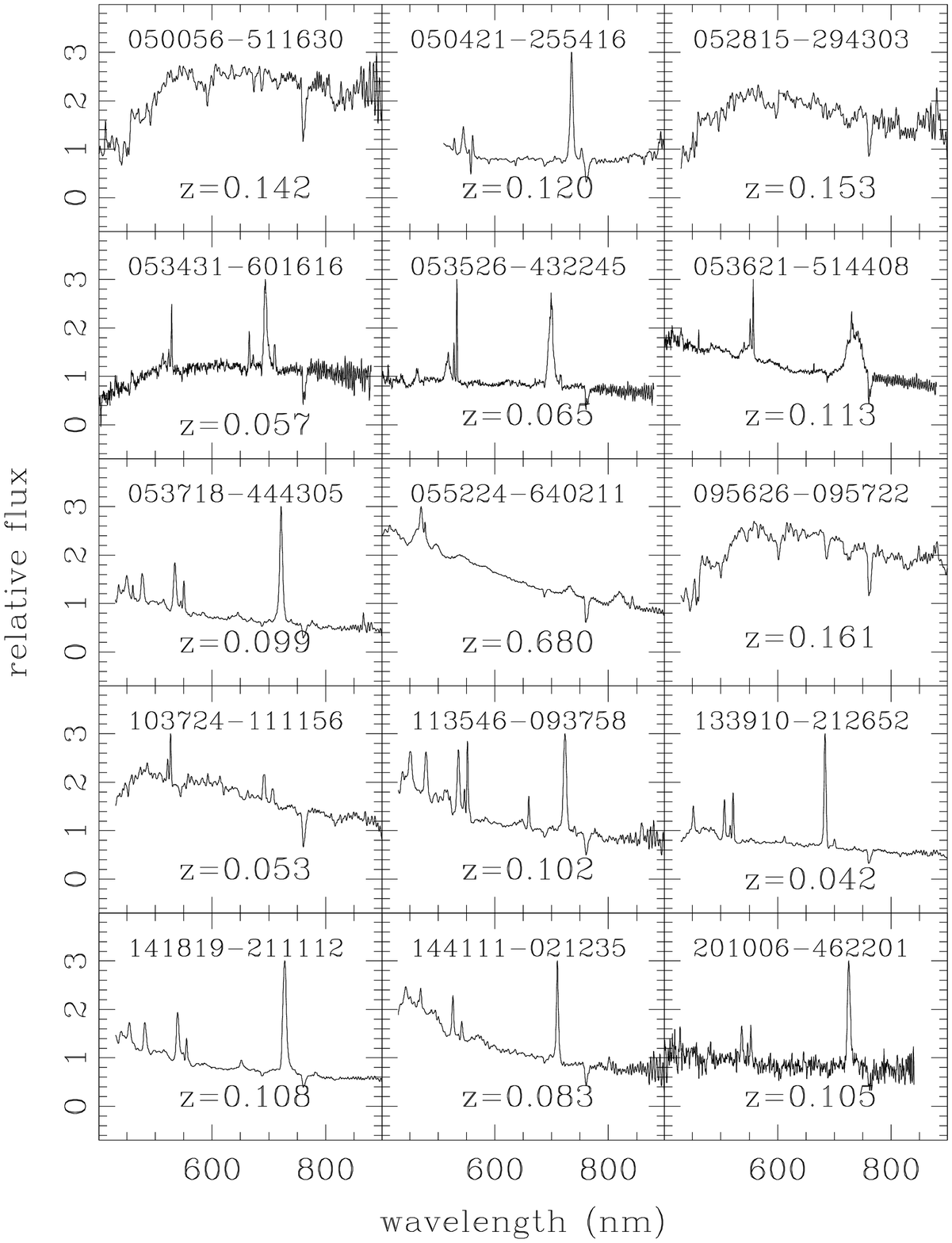}
\caption{Continued.\label{spec4}}
\end{figure}

\begin{figure}
\plotone{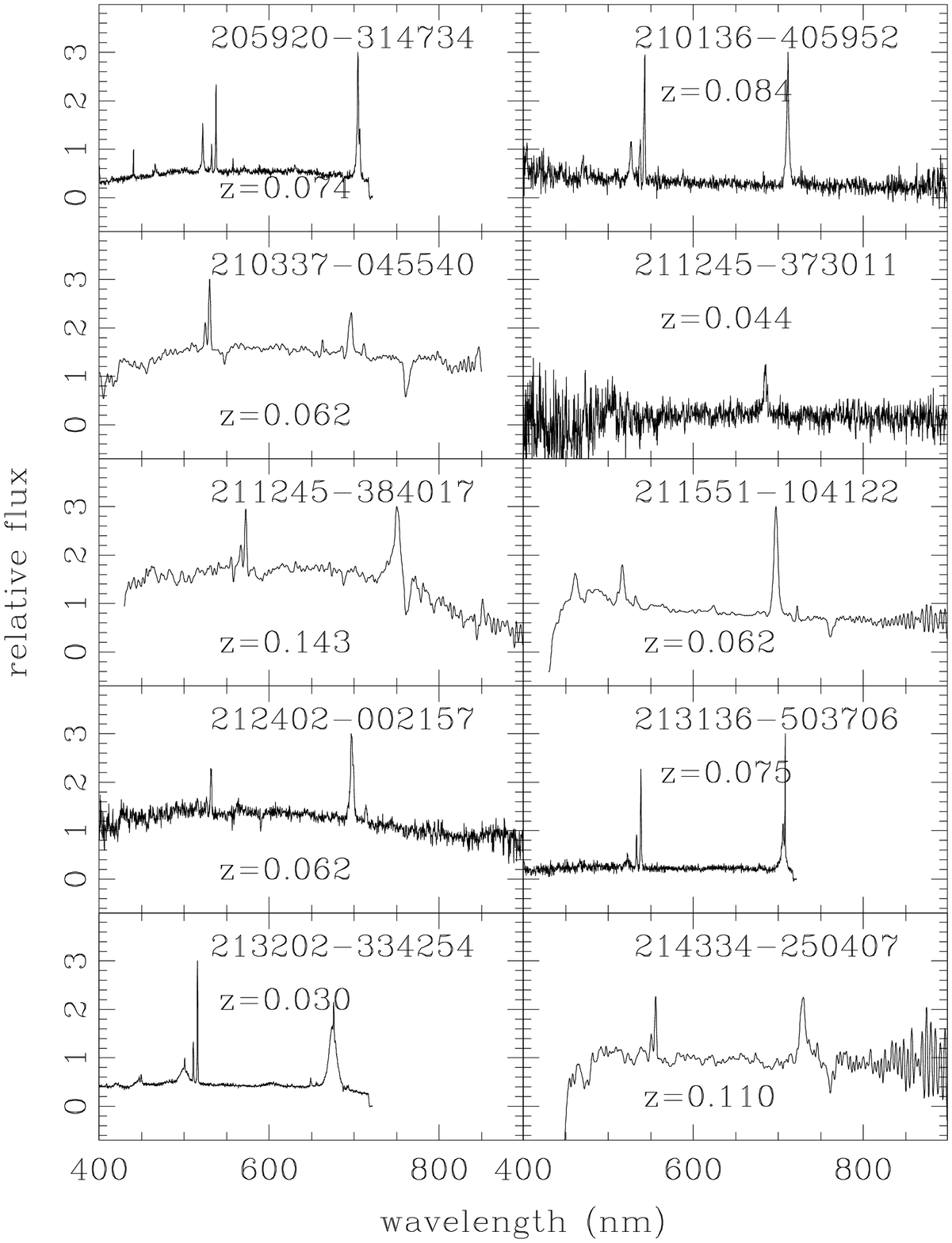}
\caption{Continued.\label{spec5}}
\end{figure}

\begin{figure}
\plotone{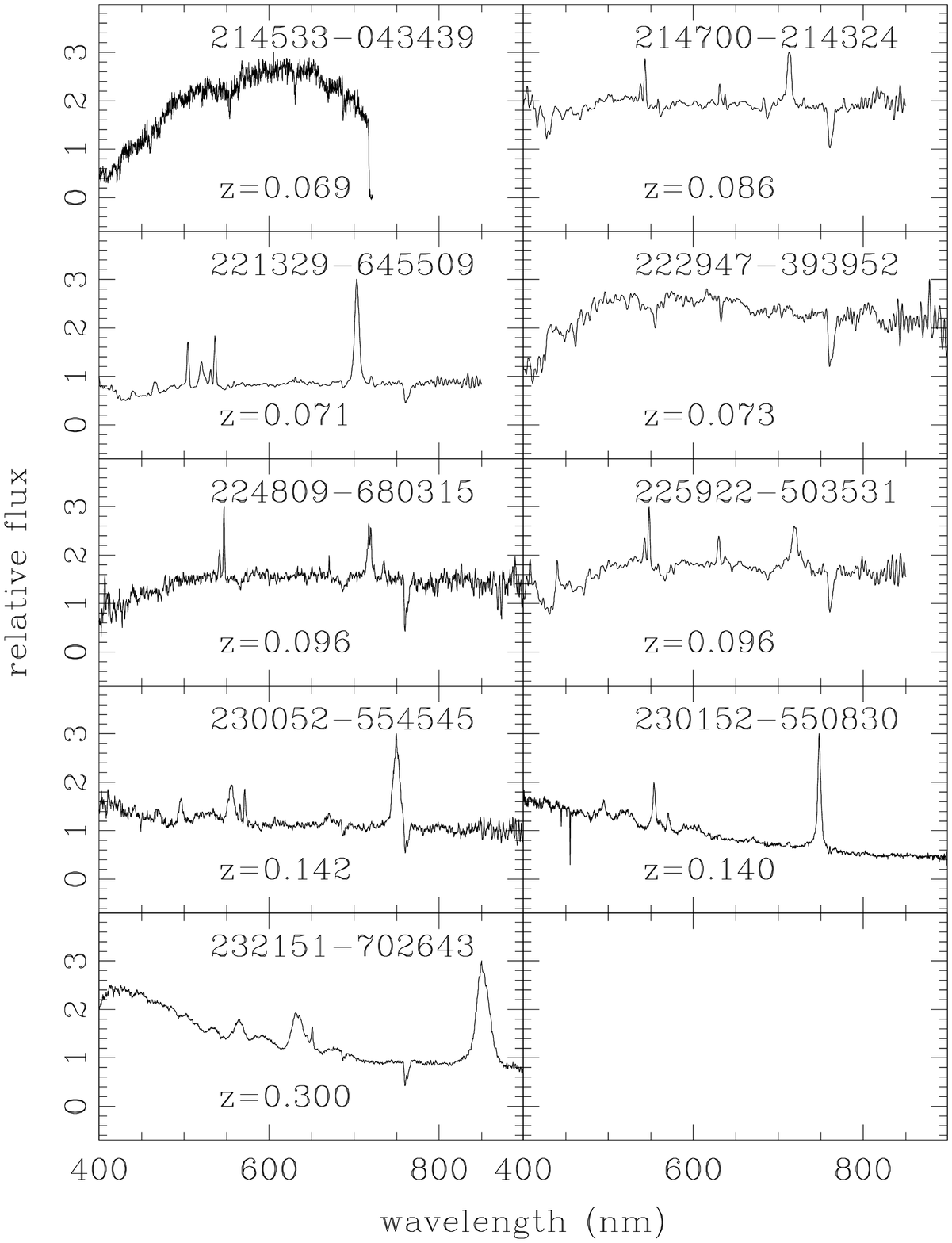}
\caption{Continued.\label{spec6}}
\end{figure}

\begin{figure}
\plotone{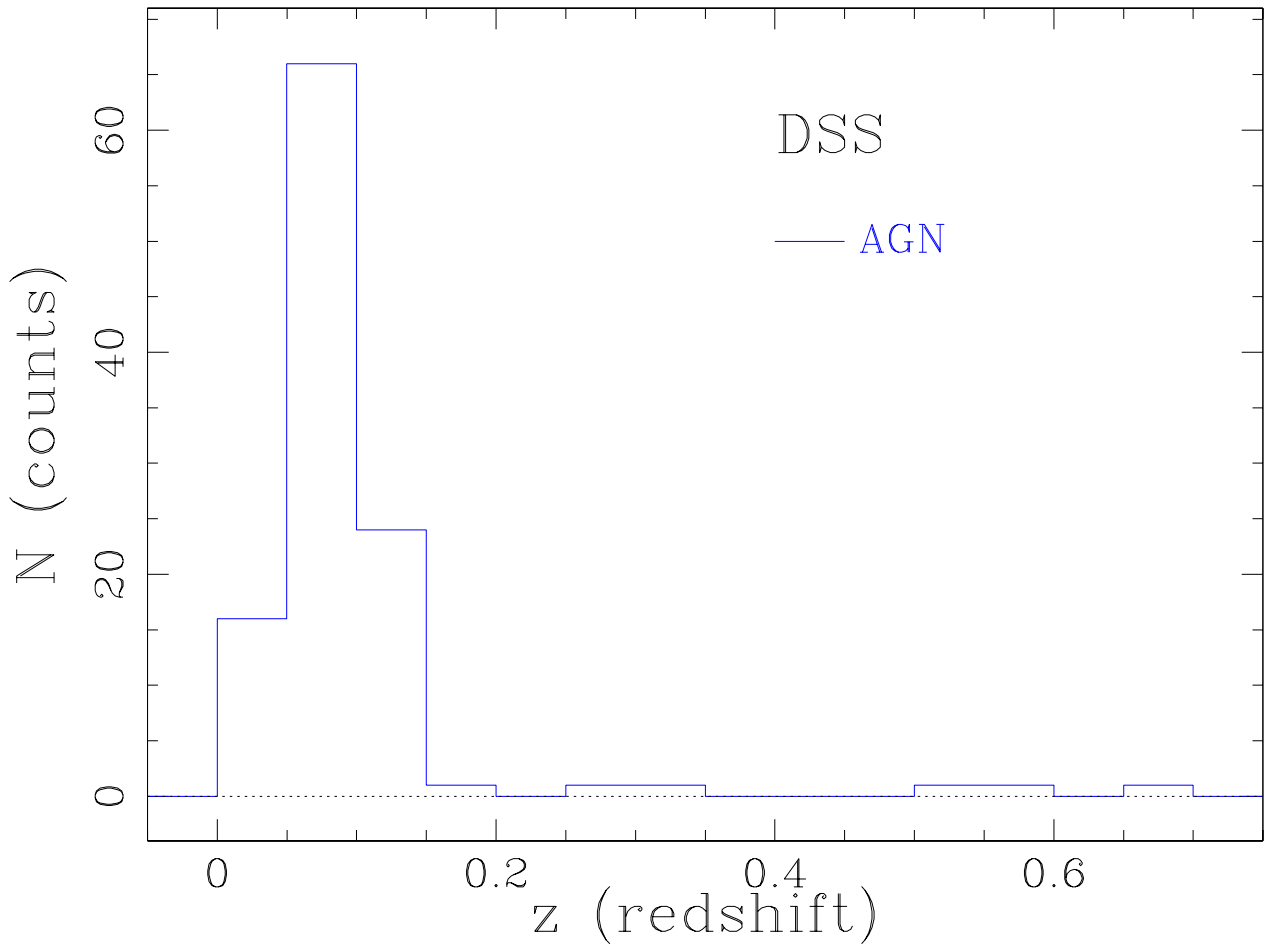}
\caption{The redshift distribution for the AGN
in our Southern Sample.\label{histz}}
\end{figure}

\begin{figure}
\plotone{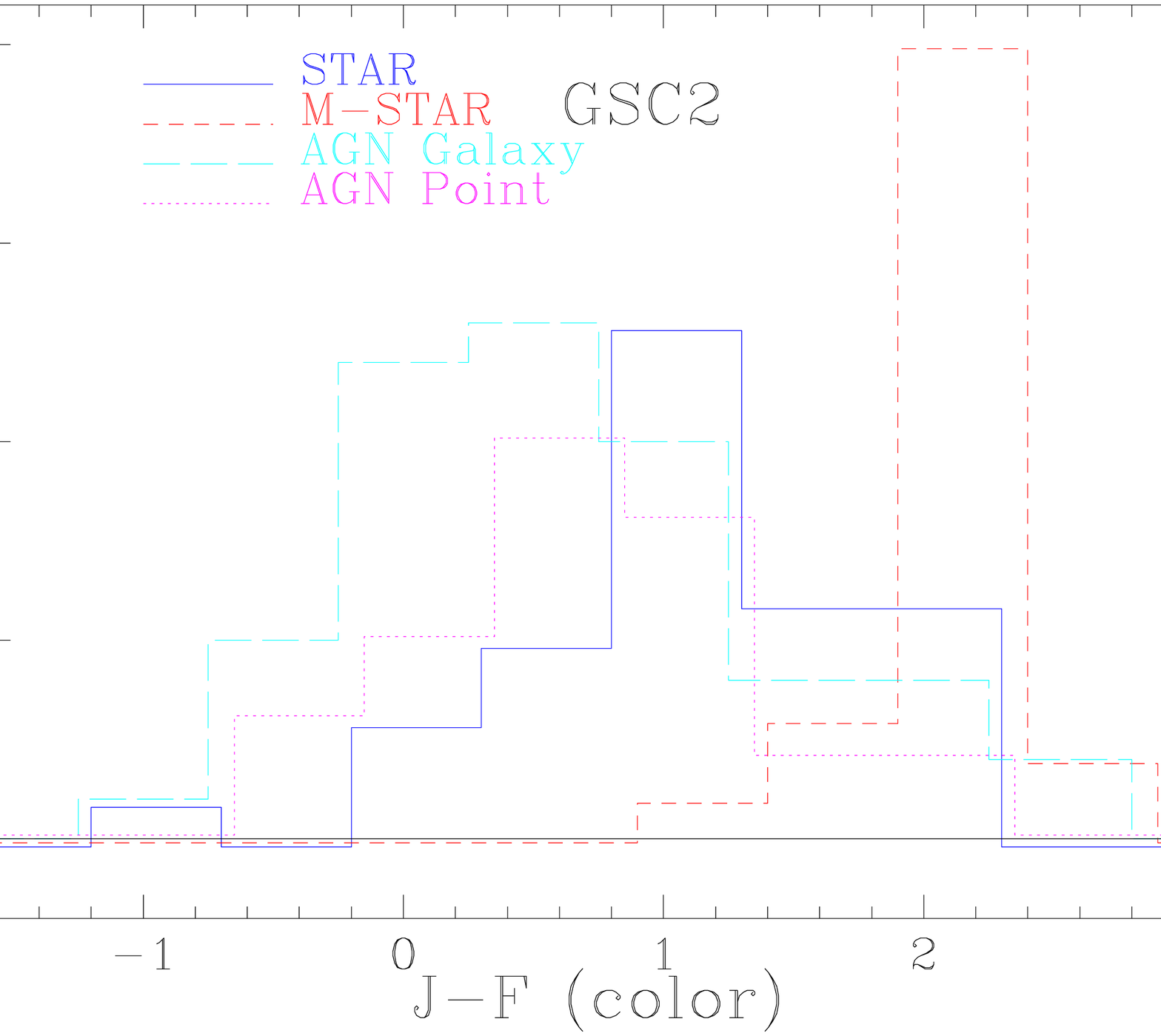}
\caption{The $J-F$ color distribution for AGN, M-stars and normal stars
in our Southern Sample. Only M-stars can be separated from AGN with
optical color
criteria. ``AGN Point'' refers to AGN classified as point like sources in the
GSC-2 catalogue. ``AGN Galaxy'' are AGN classified as extended sources.
Histograms are shifted slightly in x and y directions for
clarity.\label{histbr}}
\end{figure}

\begin{figure}
\plotone{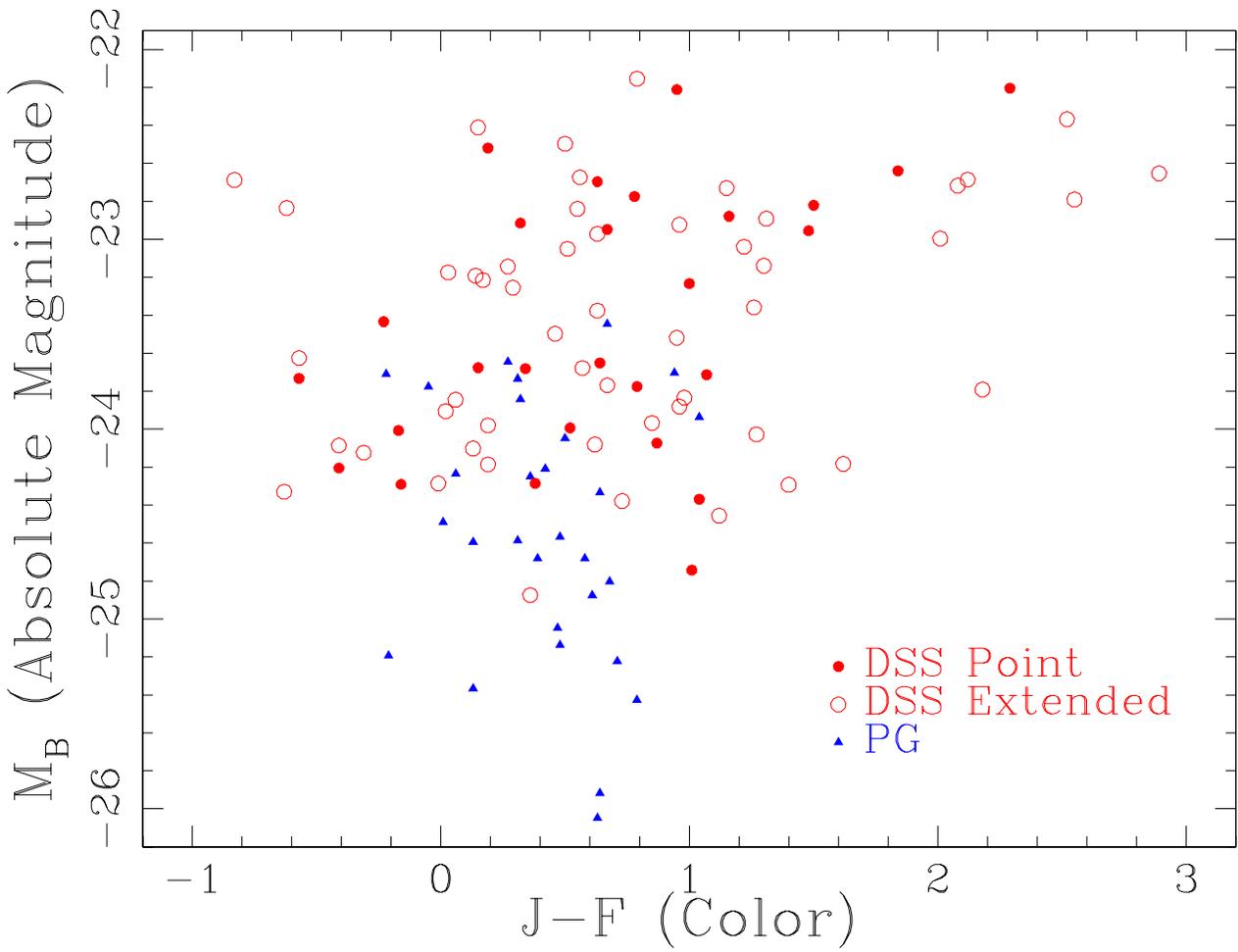}
\caption{The $J-F$ color vs. Absolute Magnitude $M_B$ for AGN in our Southern
Sample and for QSOs belonging to the PG survey \citep{PG83}.
Faint Nuclei are redder than bright QSOs: the host galaxy starts to
affect the optical color of the AGN.\label{mabsjf}}
\end{figure}

\begin{figure}
\plotone{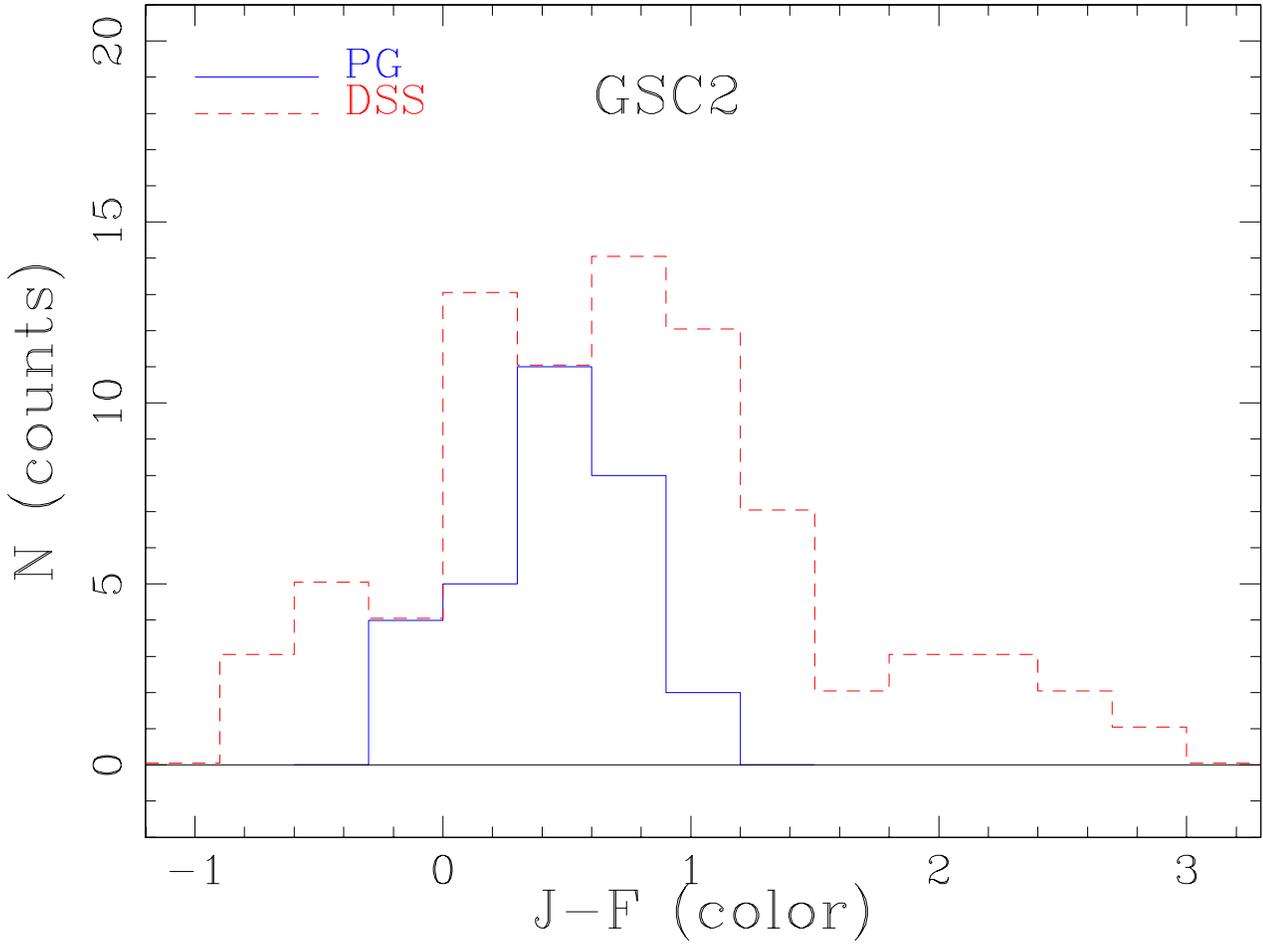}
\caption{The $J-F$ color distribution for AGN in our Southern Sample and for
QSOs belonging to the PG survey \citep{PG83}.
A simple optical color selection $J-F\le 1.0$, would decrease dramatically
the completeness by a factor of 28$\%$.
Histograms are shifted slightly in y direction for clarity.\label{histpg}}
\end{figure}

\begin{figure}
\plotone{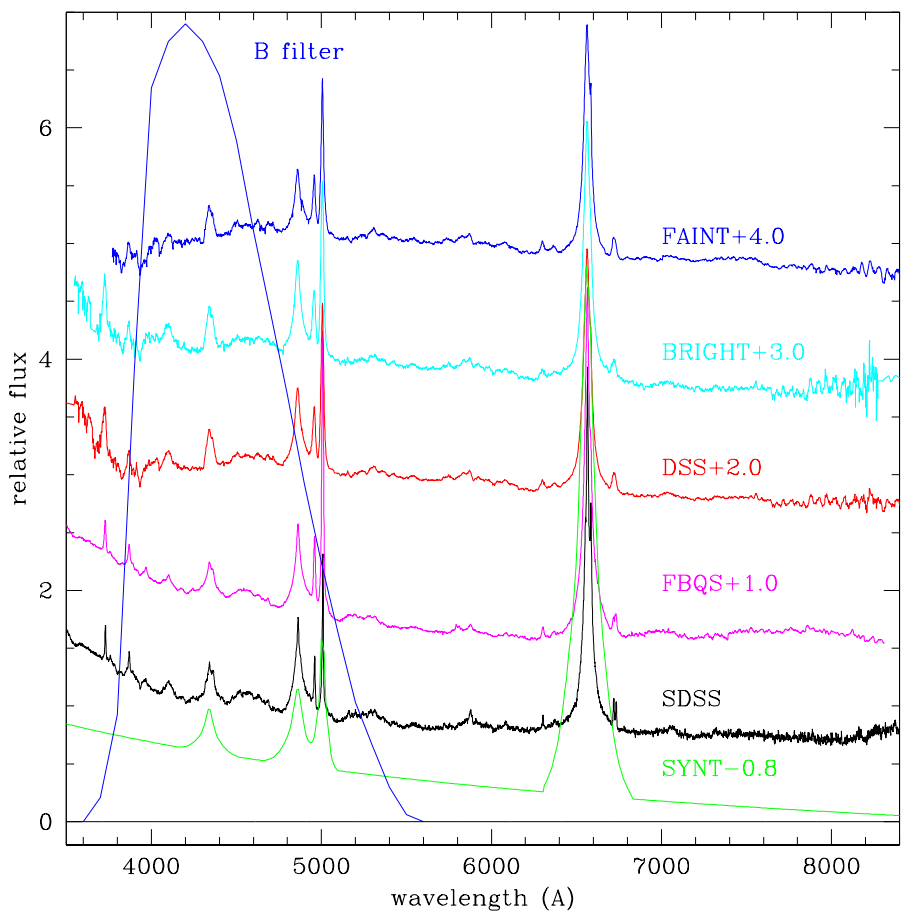}
\caption{The composite QSO spectra of this survey (DSS), of the First Bright
QSO Survey (FBQS), of the Sloan Survey (SDSS) and a synthetic spectrum (SYNT)
with $f_{\nu} \propto \nu^{-1.75}$, typically used in photometric redshift
studies. The composite spectra of bright ($M_B\le -24$, BRIGHT) QSOs in this
survey is clearly different from the composite spectra of faint ($M_B\ge -24$,
FAINT) ones. The spectra are shifted by a constant value (+4.0 for FAINT,
+3.0 for BRIGHT, +2.0 for DSS, +1.0
for FBQS and -0.8 for SYNT) for clarity.\label{composite}}
\end{figure}

\begin{figure}
\plotone{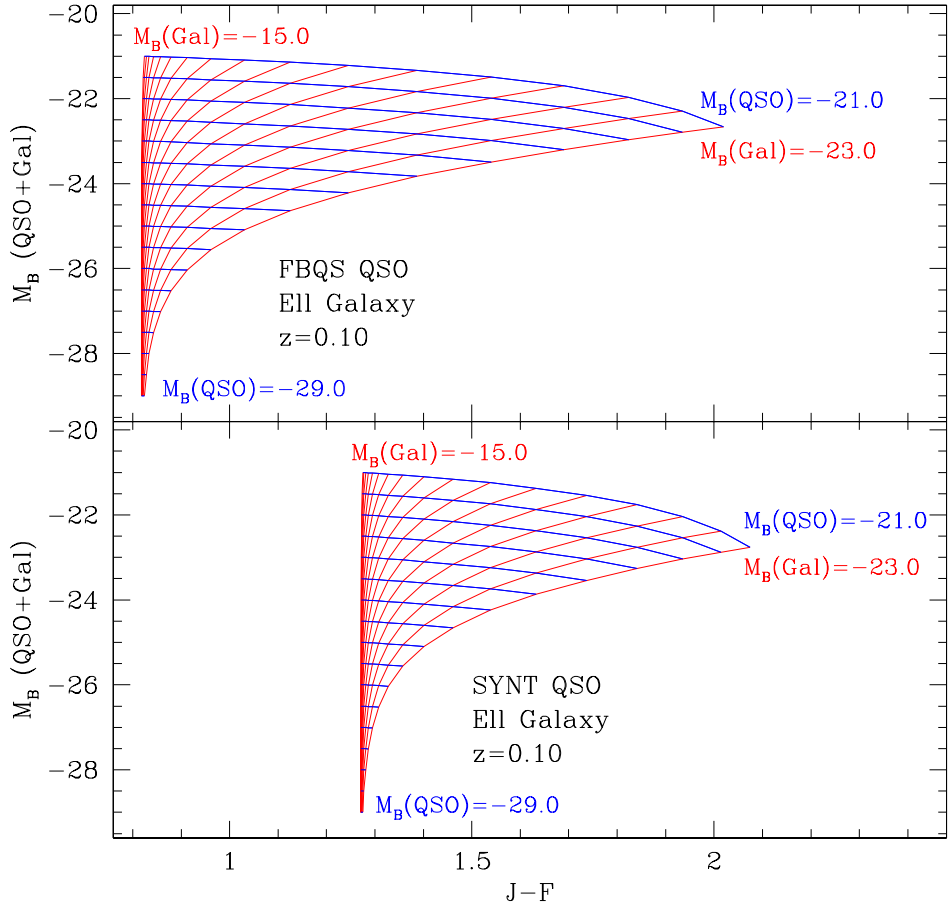}
\caption{A synthetic $J-F$ vs. $M_B$ diagram for FBQS (up) and for SYNT (down)
QSO composite spectrum
contaminated by an Elliptical galaxy, for different value of QSO and host
galaxy Absolute Magnitude. Horizontal and vertical lines represent constant
QSO and galaxy magnitudes, respectively.\label{fbqsredell}}
\end{figure}

\clearpage

\begin{deluxetable}{rrrrrr}
\tablecolumns{6}
\tablewidth{0pc}
\tablecaption{The AERQS Southern Sample.\label{dsssample}}
\tablehead{
\colhead{Name (1RXS)} & \colhead{$R.A.$} & \colhead{Declination} & \colhead{$B_J$} & \colhead{$z$} & \colhead{Type}}
\startdata
J000154.2-670749 & 00 01 55.05 & -67 07 43.43 & 14.67 & 0.0000 & STAR \\
J000307.6-180550 & 00 03 07.89 & -18 05 50.17 & 13.28 & 0.0543 & BLLAC \\
J001010.2-044225 & 00 10 10.77 & -04 42 35.39 & 14.05 & 0.0295 & AGN \\
J001020.5-061703 & 00 10 19.99 & -06 17 06.40 & 15.01 & 0.0780 & AGN \\
J001042.0-203916 & 00 10 42.65 & -20 39 03.56 & 14.14 & 0.0000 & STAR \\
J001410.1-071200 & 00 14 10.22 & -07 11 56.76 & 13.84 & 0.0000 & STAR \\
J001557.5-163659 & 00 15 58.51 & -16 36 57.42 & 14.56 & 0.0000 & STAR \\
J001936.2-071325 & 00 19 36.54 & -07 13 24.10 & 13.93 & 0.0000 & STAR \\
J002108.1-190950 & 00 21 07.53 & -19 10 05.31 & 14.86 & 0.0952 & GAL \\
J002246.1-380635 & 00 22 45.66 & -38 06 53.14 & 15.04 & 0.1190 & EM GAL \\
J002252.2-121233 & 00 22 51.51 & -12 12 31.83 & 14.76 & 0.0000 & STAR \\
J002339.6-175352 & 00 23 39.39 & -17 53 53.16 & 14.85 & 0.0535 & AGN \\
J002355.9-180254 & 00 23 55.37 & -18 02 49.53 & 14.80 & 0.0530 & AGN \\
J002750.4-323317 & 00 27 50.00 & -32 33 06.12 & 14.13 & 0.0000 & STAR \\
J003041.2-132130 & 00 30 40.18 & -13 21 29.95 & 14.56 & 0.0760 & AGN \\
J003322.1-691502 & 00 33 20.83 & -69 15 14.06 & 14.21 & 0.0977 & GAL \\
J003400.9-335428 & 00 34 01.66 & -33 54 22.07 & 15.04 & 0.1180 & AGN \\
J003908.2-222002 & 00 39 08.16 & -22 20 02.14 & 14.00 & 0.0644 & GAL \\
J004053.2-074201 & 00 40 52.75 & -07 42 09.11 & 13.05 & 0.0560 & AGN \\
J004131.9-223834 & 00 41 32.03 & -22 38 38.36 & 13.24 & 0.0630 & AGN \\
J004236.9-104919 & 00 42 36.84 & -10 49 21.93 & 13.80 & 0.0413 & AGN \\
J004423.9-261600 & 00 44 23.79 & -26 16 06.35 & 14.60 & 0.0610 & AGN \\
J004426.0-274848 & 00 44 25.39 & -27 48 57.96 & 14.57 & 0.0000 & STAR \\
J004554.8-172325 & 00 45 54.71 & -17 23 28.72 & 14.69 & 0.0970 & EM GAL \\
J005011.3-033743 & 00 50 10.62 & -03 37 53.62 & 14.62 & 0.0000 & STAR \\
J005118.0-144751 & 00 51 17.63 & -14 47 51.61 & 14.65 & 0.0910 & AGN \\
J005620.1-093626 & 00 56 20.05 & -09 36 31.10 & 13.58 & 0.1010 & BLLAC \\
J005655.1-751349 & 00 56 55.12 & -75 13 52.54 & 15.04 & 0.0740 & AGN \\
J005720.4-222300 & 00 57 20.16 & -22 22 56.50 & 13.41 & 0.0620 & AGN \\
J005822.8-024126 & 00 58 22.30 & -02 41 42.43 & 14.43 & 0.0728 & AGN \\
J010434.1-235919 & 01 04 33.90 & -23 58 29.31 & 14.85 & 0.1596 & GAL \\
J010538.7-141610 & 01 05 38.86 & -14 16 14.27 & 13.84 & 0.0670 & AGN \\
J010607.2-235907 & 01 06 07.75 & -23 59 31.52 & 14.86 & 0.0000 & STAR \\
J010818.9-413319 & 01 08 18.83 & -41 33 08.03 & 13.33 & 0.0647 & AGN \\
J010921.4-172057 & 01 09 21.69 & -17 21 03.28 & 14.22 & 0.0520 & EM GAL \\
J011029.4-151018 & 01 10 28.94 & -15 10 08.25 & 14.82 & 0.0000 & STAR \\
J011123.8-052539 & 01 11 23.55 & -05 25 39.07 & 14.68 & 0.0000 & STAR \\
J011151.3-404538 & 01 11 51.20 & -40 45 44.25 & 14.18 & 0.0540 & AGN \\
J011350.0-145041 & 01 13 50.04 & -14 50 46.46 & 13.13 & 0.0527 & AGN \\
J011457.6-422445 & 01 14 57.65 & -42 24 49.50 & 14.72 & 0.1240 & AGN \\
J011501.3-340008 & 01 15 01.47 & -33 59 26.88 & 13.98 & 0.0000 & STAR \\
J011724.1-222748 & 01 17 24.37 & -22 27 59.97 & 14.70 & 0.1180 & EM GAL \\
J011811.6-265819 & 01 18 10.63 & -26 58 46.81 & 14.71 & 0.0000 & STAR \\
J012020.1-102510 & 01 20 18.81 & -10 25 30.40 & 14.90 & 0.0000 & STAR \\
J012021.9-051052 & 01 20 21.97 & -05 10 48.18 & 14.99 & 0.0470 & GAL \\
J012059.4-270133 & 01 20 58.47 & -27 01 44.29 & 13.90 & 0.0539 & GAL \\
J012149.3-135810 & 01 21 49.95 & -13 58 10.02 & 14.66 & 0.0550 & AGN \\
J012151.5-282048 & 01 21 51.53 & -28 20 57.34 & 14.39 & 0.1170 & AGN \\
J012250.4-243937 & 01 22 50.49 & -24 39 44.35 & 15.12 & 0.0000 & STAR \\
J012448.3-115823 & 01 24 48.30 & -11 58 08.87 & 14.93 & 0.0680 & AGN \\
J012749.6-265036 & 01 27 50.17 & -26 50 40.85 & 14.92 & 0.1090 & AGN \\
J012806.9-184837 & 01 28 06.71 & -18 48 31.10 & 13.14 & 0.0430 & AGN \\
J013020.0-255710 & 01 30 20.41 & -25 57 10.69 & 14.52 & 0.0000 & STAR \\
J013445.2-043017 & 01 34 45.65 & -04 30 13.61 & 14.87 & 0.0790 & AGN \\
J013449.4-025441 & 01 34 50.33 & -02 54 41.29 & 14.36 & 0.0000 & STAR \\
J013514.2-071254 & 01 35 13.61 & -07 12 49.72 & 14.19 & 0.0000 & STAR \\
J013635.8-080617 & 01 36 35.53 & -08 06 06.87 & 15.11 & 0.1461 & GAL \\
J013655.2-064731 & 01 36 54.62 & -06 47 34.04 & 15.03 & 0.0000 & STAR \\
J014132.8-152755 & 01 41 32.53 & -15 28 01.88 & 13.64 & 0.0820 & AGN \\
J014345.1-060239 & 01 43 44.93 & -06 02 39.34 & 14.04 & 0.0000 & STAR \\
J014442.0-221339 & 01 44 40.43 & -22 13 46.60 & 14.52 & 0.2780 & GAL \\
J014841.1-483057 & 01 48 40.62 & -48 30 51.48 & 13.78 & 0.0000 & STAR \\
J015211.3-210737 & 01 52 11.34 & -21 07 42.46 & 14.82 & 0.1040 & EM GAL \\
J015227.1-231956 & 01 52 27.06 & -23 19 53.90 & 14.50 & 0.1130 & AGN \\
J015440.5-270659 & 01 54 40.26 & -27 07 00.52 & 14.83 & 0.1510 & AGN \\
J015503.5-050835 & 01 55 02.96 & -05 08 34.55 & 15.10 & 0.1290 & AGN \\
J015948.9-035206 & 01 59 49.04 & -03 52 00.34 & 15.12 & 0.0000 & STAR \\
J020013.6-084106 & 02 00 12.39 & -08 40 48.90 & 13.46 & 0.0000 & STAR \\
J020058.2-621451 & 02 01 01.48 & -62 14 34.18 & 14.92 & 0.0000 & STAR \\
J020515.9-450100 & 02 05 16.47 & -45 01 02.79 & 14.90 & 0.1192 & GAL \\
J020952.1-631838 & 02 09 50.73 & -63 18 39.92 & 14.69 & 0.0000 & STAR \\
J020953.8-135321 & 02 09 53.77 & -13 53 20.87 & 13.88 & 0.0730 & AGN \\
J021125.9-401702 & 02 11 24.82 & -40 17 27.45 & 14.50 & 0.1050 & GAL \\
J021220.1-444045 & 02 12 19.04 & -44 41 05.54 & 15.03 & 0.0000 & STAR \\
J021411.4-473241 & 02 14 11.86 & -47 32 53.95 & 15.07 & 0.0000 & STAR \\
J021438.0-643018 & 02 14 36.49 & -64 30 17.50 & 13.79 & 0.0000 & STAR \\
J021559.9-092913 & 02 15 58.64 & -09 29 09.92 & 13.22 & 0.0000 & STAR \\
J021738.8-300455 & 02 17 38.15 & -30 04 48.29 & 14.99 & 0.0800 & AGN \\
J022039.7-263441 & 02 20 41.84 & -26 34 47.24 & 15.12 & 0.0000 & STAR \\
J022225.7-411553 & 02 22 25.14 & -41 15 52.27 & 14.83 & 0.0680 & GAL \\
J022742.2-335351 & 02 27 42.34 & -33 53 48.88 & 14.12 & 0.0000 & STAR \\
J022901.8-153856 & 02 29 01.71 & -15 38 54.10 & 14.88 & 0.0590 & AGN \\
J023343.2-221744 & 02 33 45.11 & -22 17 44.20 & 14.87 & 0.0000 & STAR \\
J023400.1-181155 & 02 33 59.64 & -18 11 51.90 & 14.83 & 0.0000 & STAR \\
J023434.1-520359 & 02 34 34.31 & -52 03 55.26 & 14.79 & 0.1370 & AGN \\
J023849.4-403844 & 02 38 48.90 & -40 38 39.05 & 13.18 & 0.0620 & AGN \\
J024115.7-480733 & 02 41 17.34 & -48 07 37.02 & 14.64 & 0.0000 & STAR \\
J024146.8-525943 & 02 41 47.12 & -52 59 30.19 & 12.86 & 0.0000 & STAR \\
J024515.7-462754 & 02 45 13.36 & -46 27 19.70 & 14.62 & 0.0920 & GAL \\
J024554.2-445942 & 02 45 51.83 & -44 59 44.95 & 15.09 & 0.0000 & STAR \\
J024853.5-340428 & 02 48 52.45 & -34 04 25.72 & 14.61 & 0.0000 & STAR \\
J025126.1-245653 & 02 51 24.83 & -24 56 39.51 & 14.18 & 0.1130 & GAL \\
J025407.6-413731 & 02 54 07.04 & -41 37 32.44 & 14.76 & 0.1460 & AGN \\
J031521.0-564246 & 03 15 21.35 & -56 42 51.19 & 14.91 & 0.0730 & AGN \\
J031920.9-414639 & 03 19 20.22 & -41 46 39.04 & 14.40 & 0.0810 & AGN \\
J032214.3-664714 & 03 22 11.55 & -66 47 28.86 & 15.01 & 0.0980 & GAL \\
J032315.7-493113 & 03 23 15.28 & -49 31 06.38 & 13.66 & 0.0710 & AGN \\
J032521.8-563543 & 03 25 23.58 & -56 35 45.45 & 13.97 & 0.0610 & GAL \\
J033307.5-135419 & 03 33 07.77 & -13 54 33.19 & 13.80 & 0.0390 & AGN \\
J033424.5-151325 & 03 34 24.53 & -15 13 40.69 & 13.45 & 0.0350 & AGN \\
J033451.2-534242 & 03 34 51.53 & -53 42 38.19 & 14.94 & 0.0613 & GAL \\
J033648.2-554519 & 03 36 47.75 & -55 45 12.61 & 15.03 & 0.0000 & STAR \\
J033807.3-553558 & 03 38 06.27 & -55 36 00.39 & 13.27 & 0.0590 & AGN \\
J033823.5-451057 & 03 38 23.24 & -45 10 49.22 & 14.97 & 0.1190 & AGN \\
J034039.1-524301 & 03 40 38.35 & -52 42 59.55 & 14.75 & 0.0000 & STAR \\
J034117.1-225228 & 03 41 15.93 & -22 52 43.14 & 14.13 & 0.0000 & STAR \\
J034716.3-044419 & 03 47 16.34 & -04 44 15.86 & 14.90 & 0.0000 & STAR \\
J034930.8-534439 & 03 49 32.40 & -53 44 09.09 & 14.72 & 0.0000 & STAR \\
J035432.5-134005 & 03 54 32.81 & -13 40 08.33 & 15.09 & 0.0766 & AGN \\
J040126.6-080143 & 04 01 26.30 & -08 01 59.92 & 14.59 & 0.1470 & AGN ? \\
J040748.7-121133 & 04 07 48.42 & -12 11 36.67 & 14.64 & 0.5740 & AGN \\
J040805.1-273136 & 04 08 05.48 & -27 31 38.31 & 14.55 & 0.0000 & STAR \\
J040913.8-112455 & 04 09 13.51 & -11 25 02.43 & 14.58 & 0.0920 & AGN \\
J041417.0-090650 & 04 14 16.93 & -09 06 48.82 & 14.45 & 0.0000 & STAR \\
J041420.6-594134 & 04 14 19.05 & -59 41 32.14 & 15.02 & 0.0710 & AGN \\
J041530.5-661937 & 04 15 30.42 & -66 19 19.85 & 14.66 & 0.0000 & STAR \\
J041756.9-382649 & 04 17 57.33 & -38 27 02.80 & 14.51 & 0.0000 & STAR \\
J042202.2-415324 & 04 22 01.90 & -41 53 28.86 & 14.13 & 0.0621 & AGN \\
J042947.7-305240 & 04 29 43.69 & -30 52 54.30 & 14.21 & 0.0000 & STAR \\
J043153.6-585218 & 04 31 50.31 & -58 52 12.17 & 14.86 & 0.0000 & STAR \\
J043520.2-780150 & 04 35 16.29 & -78 01 56.59 & 13.05 & 0.0610 & AGN \\
J043726.6-471118 & 04 37 28.08 & -47 11 29.43 & 13.97 & 0.0520 & AGN \\
J044154.5-082639 & 04 41 54.00 & -08 26 34.33 & 14.49 & 0.0440 & AGN \\
J044404.7-222441 & 04 44 03.94 & -22 24 46.30 & 14.94 & 0.0760 & AGN \\
J044708.2-265731 & 04 47 07.78 & -26 57 44.24 & 14.97 & 0.0000 & STAR \\
J045230.4-295329 & 04 52 30.05 & -29 53 35.20 & 15.04 & 0.2860 & AGN \\
J045816.3-751608 & 04 58 17.19 & -75 16 10.92 & 15.12 & 0.0000 & STAR \\
J045851.2-190542 & 04 58 50.60 & -19 06 04.32 & 14.52 & 0.0620 & AGN \\
J045958.1-611506 & 04 59 57.74 & -61 15 10.15 & 13.61 & 0.0860 & AGN \\
J050054.7-511547 & 05 00 56.84 & -51 16 30.68 & 13.97 & 0.1420 & GAL \\
J050421.9-255420 & 05 04 22.05 & -25 54 16.13 & 15.00 & 0.1200 & AGN \\
J050903.4-420926 & 05 09 03.39 & -42 09 21.92 & 14.64 & 0.0000 & STAR \\
J051004.7-234024 & 05 10 04.17 & -23 40 40.73 & 14.11 & 0.0000 & STAR \\
J051949.5-454644 & 05 19 49.64 & -45 46 44.15 & 13.70 & 0.0351 & AGN \\
J052258.0-362729 & 05 22 57.96 & -36 27 31.35 & 13.14 & 0.0553 & AGN \\
J052815.9-294305 & 05 28 15.08 & -29 43 03.04 & 14.95 & 0.1530 & GAL \\
J052925.8-324858 & 05 29 25.38 & -32 49 01.31 & 13.47 & 0.0000 & STAR \\
J052945.2-323911 & 05 29 44.64 & -32 39 14.58 & 15.11 & 0.0000 & STAR \\
J053431.8-601613 & 05 34 31.03 & -60 16 16.03 & 14.46 & 0.0570 & AGN \\
J053509.9-390557 & 05 35 12.51 & -39 06 05.65 & 14.83 & 0.0000 & STAR \\
J053527.5-432247 & 05 35 26.78 & -43 22 45.83 & 14.08 & 0.0650 & AGN \\
J053555.0-653039 & 05 35 54.65 & -65 30 38.67 & 14.80 & 0.0000 & STAR \\
J053602.5-471844 & 05 36 02.87 & -47 18 49.79 & 14.14 & 0.0000 & STAR \\
J053621.3-514401 & 05 36 21.23 & -51 44 08.20 & 14.69 & 0.1130 & AGN \\
J053718.6-444257 & 05 37 18.66 & -44 43 05.01 & 14.19 & 0.0990 & AGN \\
J054105.5-615122 & 05 41 04.42 & -61 51 50.68 & 14.99 & 0.0000 & STAR \\
J055225.0-640206 & 05 52 24.54 & -64 02 11.37 & 15.04 & 0.6800 & AGN \\
J093444.7-060930 & 09 34 44.95 & -06 09 19.44 & 13.97 & 0.0000 & STAR \\
J095627.2-095720 & 09 56 26.41 & -09 57 22.36 & 14.81 & 0.1610 & GAL \\
J100802.7-145904 & 10 08 02.81 & -14 59 05.93 & 13.57 & 0.0560 & AGN \\
J100816.5-031526 & 10 08 16.60 & -03 15 31.25 & 14.75 & 0.0000 & STAR \\
J101438.9-084450 & 10 14 38.90 & -08 45 20.27 & 14.99 & 0.0000 & STAR \\
J101907.1-053703 & 10 19 07.26 & -05 37 13.40 & 14.70 & 0.0747 & AGN \\
J102225.1-142859 & 10 22 24.80 & -14 28 57.61 & 15.05 & 0.0770 & AGN \\
J102758.9-064804 & 10 27 58.68 & -06 47 56.76 & 14.35 & 0.1165 & AGN \\
J103727.1-111124 & 10 37 24.33 & -11 11 56.00 & 14.42 & 0.0530 & AGN \\
J103743.2-054848 & 10 37 43.85 & -05 48 55.66 & 14.79 & 0.0000 & STAR \\
J104115.4-210124 & 10 41 15.10 & -21 01 25.21 & 13.26 & 0.0120 & AGN \\
J104617.3-140206 & 10 46 17.08 & -14 02 27.75 & 14.92 & 0.0680 & AGN \\
J105421.2-092154 & 10 54 20.83 & -09 21 56.52 & 13.62 & 0.0630 & AGN \\
J112913.4-172114 & 11 29 14.18 & -17 21 17.64 & 14.67 & 0.0000 & STAR \\
J113104.6-094353 & 11 31 05.06 & -09 43 53.72 & 14.55 & 0.0000 & STAR \\
J113241.7-265155 & 11 32 41.50 & -26 51 54.79 & 13.11 & 0.0000 & STAR \\
J113301.6-153153 & 11 33 00.24 & -15 31 51.56 & 14.60 & 0.0000 & STAR \\
J113526.8-284040 & 11 35 26.14 & -28 40 37.70 & 14.88 & 0.0820 & AGN \\
J113546.6-093748 & 11 35 46.18 & -09 37 58.37 & 14.91 & 0.1020 & AGN \\
J113923.1-083241 & 11 39 21.93 & -08 32 28.39 & 14.56 & 0.0000 & STAR \\
J114042.0-174008 & 11 40 42.22 & -17 40 10.38 & 12.94 & 0.0210 & AGN \\
J114918.8-041649 & 11 49 18.64 & -04 16 51.42 & 14.58 & 0.0850 & AGN \\
J120246.0-034710 & 12 02 45.33 & -03 47 21.48 & 14.49 & 0.0645 & AGN \\
J120622.6-131453 & 12 06 21.90 & -13 14 53.23 & 13.77 & 0.0000 & STAR \\
J121027.7-131029 & 12 10 27.60 & -13 10 08.51 & 14.87 & 0.0000 & STAR \\
J131231.3-322847 & 13 12 30.72 & -32 28 45.62 & 14.92 & 0.0000 & STAR \\
J133910.9-212650 & 13 39 10.88 & -21 26 52.15 & 14.39 & 0.0420 & AGN \\
J134209.9-160020 & 13 42 11.40 & -16 00 22.14 & 14.70 & 0.0000 & STAR \\
J134951.0-131338 & 13 49 51.89 & -13 13 38.03 & 15.05 & 0.0000 & STAR \\
J135734.0-125433 & 13 57 33.20 & -12 54 18.61 & 14.66 & 0.0581 & GAL \\
J140329.8-084018 & 14 03 28.96 & -08 40 23.84 & 14.92 & 0.0890 & AGN \\
J141632.9-072529 & 14 16 33.15 & -07 25 37.09 & 14.91 & 0.0000 & STAR \\
J141817.0-211048 & 14 18 19.38 & -21 11 12.01 & 14.13 & 0.1080 & AGN \\
J142342.3-151550 & 14 23 42.03 & -15 15 55.01 & 14.73 & 0.0000 & STAR \\
J144111.3-021225 & 14 41 11.52 & -02 12 35.28 & 14.94 & 0.0830 & AGN \\
J144327.5-162029 & 14 43 30.09 & -16 20 33.00 & 14.63 & 0.0000 & STAR \\
J144427.6-042410 & 14 44 27.75 & -04 24 03.60 & 14.77 & 0.0000 & STAR \\
J150957.2-022554 & 15 09 57.82 & -02 26 03.34 & 14.71 & 0.0000 & STAR \\
J155542.1-102012 & 15 55 42.04 & -10 20 00.09 & 14.68 & 0.0000 & STAR \\
J201006.7-462206 & 20 10 06.86 & -46 22 01.45 & 14.55 & 0.1050 & AGN \\
J204644.0-114803 & 20 46 42.58 & -11 48 10.84 & 14.84 & 0.0000 & STAR \\
J205920.9-314733 & 20 59 20.72 & -31 47 35.27 & 14.20 & 0.0740 & AGN \\
J210134.8-410005 & 21 01 35.99 & -40 59 51.94 & 14.35 & 0.0840 & AGN \\
J210338.0-045548 & 21 03 37.89 & -04 55 40.40 & 14.56 & 0.0620 & AGN \\
J210736.5-130500 & 21 07 36.61 & -13 04 54.44 & 13.75 & 0.0000 & STAR \\
J210759.4-375400 & 21 07 59.77 & -37 54 09.65 & 14.26 & 0.0490 & AGN \\
J210910.1-094011 & 21 09 08.88 & -09 40 18.55 & 12.65 & 0.0270 & AGN \\
J211208.1-085004 & 21 12 11.17 & -08 49 58.58 & 14.24 & 0.0000 & STAR \\
J211244.3-373019 & 21 12 44.84 & -37 30 12.24 & 13.77 & 0.0440 & AGN \\
J211245.4-384025 & 21 12 45.09 & -38 40 17.12 & 14.97 & 0.1430 & AGN \\
J211551.4-104109 & 21 15 51.26 & -10 41 22.27 & 14.51 & 0.0620 & AGN \\
J212352.8-390819 & 21 23 52.79 & -39 08 17.09 & 14.61 & 0.0000 & STAR \\
J212401.9-002150 & 21 24 01.88 & -00 21 58.46 & 14.77 & 0.0620 & AGN \\
J212610.1-361813 & 21 26 07.60 & -36 18 45.62 & 14.85 & 0.0000 & STAR \\
J212951.7-022008 & 21 29 51.73 & -02 20 06.04 & 14.52 & 0.0000 & STAR \\
J213136.7-503704 & 21 31 36.15 & -50 37 06.71 & 14.58 & 0.0750 & AGN \\
J213135.7-120719 & 21 31 37.03 & -12 07 24.64 & 14.11 & 0.5010 & AGN \\
J213202.3-334255 & 21 32 02.25 & -33 42 54.54 & 14.26 & 0.0300 & AGN \\
J213623.1-622400 & 21 36 23.20 & -62 24 00.47 & 13.54 & 0.0590 & AGN \\
J213648.0-012407 & 21 36 49.40 & -01 24 08.21 & 14.95 & 0.0000 & STAR \\
J213704.1-340132 & 21 37 03.53 & -34 01 05.41 & 15.06 & 0.0900 & GAL \\
J214055.0-512516 & 21 40 54.17 & -51 25 20.54 & 13.99 & 0.0970 & AGN \\
J214334.0-250403 & 21 43 34.64 & -25 04 07.47 & 15.03 & 0.1100 & AGN \\
J214533.6-043434 & 21 45 33.39 & -04 34 39.43 & 13.62 & 0.0690 & GAL \\
J214701.2-214343 & 21 47 00.22 & -21 43 24.49 & 14.72 & 0.0860 & AGN \\
J215526.2-121025 & 21 55 27.79 & -12 10 05.56 & 14.61 & 0.0000 & STAR \\
J215830.1-094759 & 21 58 28.93 & -09 47 49.81 & 14.08 & 0.0803 & GAL \\
J220226.6-165755 & 22 02 26.47 & -16 57 50.58 & 14.78 & 0.0000 & STAR \\
J221142.4-204406 & 22 11 41.60 & -20 44 15.11 & 14.81 & 0.0000 & STAR \\
J221329.3-645512 & 22 13 29.53 & -64 55 09.69 & 14.51 & 0.0710 & AGN \\
J221504.1-033512 & 22 15 04.08 & -03 35 26.66 & 14.82 & 0.0000 & STAR \\
J221839.1-532639 & 22 18 40.42 & -53 26 41.31 & 13.87 & 0.0000 & STAR \\
J221959.6-505249 & 22 19 57.95 & -50 53 04.13 & 14.98 & 0.0000 & STAR \\
J223039.2-394246 & 22 29 47.72 & -39 39 52.60 & 14.90 & 0.0730 & GAL \\
J223046.8-423910 & 22 30 45.28 & -42 38 52.01 & 14.98 & 0.0000 & STAR \\
J223244.3-413441 & 22 32 43.16 & -41 34 37.13 & 14.51 & 0.0750 & AGN \\
J223455.4-605216 & 22 34 54.73 & -60 52 10.60 & 14.28 & 0.0000 & STAR \\
J224811.4-680322 & 22 48 09.31 & -68 03 14.73 & 14.63 & 0.0960 & AGN \\
J224841.4-510951 & 22 48 41.11 & -51 09 53.43 & 14.69 & 0.1000 & AGN \\
J225518.1-031040 & 22 55 17.93 & -03 10 39.58 & 12.97 & 0.0000 & STAR \\
J225923.7-503530 & 22 59 22.72 & -50 35 31.75 & 14.17 & 0.0960 & AGN \\
J230050.7-554549 & 23 00 52.03 & -55 45 45.14 & 15.05 & 0.1420 & AGN \\
J230152.0-550827 & 23 01 52.01 & -55 08 30.91 & 14.84 & 0.1400 & AGN \\
J230358.7-551717 & 23 03 57.97 & -55 17 17.59 & 15.11 & 0.0840 & AGN \\
J232046.5-672317 & 23 20 46.82 & -67 23 18.97 & 14.65 & 0.0000 & STAR \\
J232152.0-702645 & 23 21 51.16 & -70 26 43.54 & 14.97 & 0.3000 & AGN \\
J232857.5-680225 & 23 28 57.38 & -68 02 32.49 & 13.94 & 0.0000 & STAR \\
J233355.5-234336 & 23 33 55.23 & -23 43 40.47 & 13.78 & 0.0480 & AGN \\
J234032.5-263323 & 23 40 32.04 & -26 33 19.37 & 12.89 & 0.0496 & AGN \\
J234524.5-712645 & 23 45 21.95 & -71 26 49.09 & 14.91 & 0.0000 & STAR \\
J234842.8-735746 & 23 48 35.10 & -73 57 33.99 & 14.36 & 0.0000 & STAR \\
J234923.9-312602 & 23 49 23.94 & -31 26 02.98 & 14.69 & 0.1350 & AGN \\
J235555.3-132126 & 23 55 54.15 & -13 21 24.80 & 14.67 & 0.0000 & STAR \\
J235622.3-042949 & 23 56 19.77 & -04 29 31.34 & 14.79 & 0.0000 & STAR \\
J235720.0-125852 & 23 57 19.92 & -12 58 49.98 & 14.27 & 0.0000 & STAR \\
J235812.9-172437 & 23 58 12.97 & -17 24 35.17 & 12.84 & 0.0000 & STAR \\
\enddata
\tablecomments{R.A. is in $HH^h MM^m SS^s.SS$.
Declination is in  $DD^{\circ} PP^{'} SS^{``}.SS$.
The coordinate system used is J2000.
The classification and the redshift of J040126.6-080143 is uncertain, due to
the low S/N of the spectrum.}
\end{deluxetable}

\clearpage

\begin{deluxetable}{ccccccc}
\tablecolumns{7}
\tablewidth{0pc}
\tablecaption{Area covered by AERQS. [The complete version of this table
is in the electronic edition of the Journal.
The printed edition contains only a sample.]\label{areasud}}
\tablehead{
\colhead{$RA$} & \colhead{$DEC$} & \colhead{$l_{gal}$} & \colhead{$b_{gal}$} & \colhead{$Expt$} & \colhead{Plate ID} & \colhead{Area}}
\startdata
0.008333 & -0.125000 & 96.477112 & -60.35298 & 370.624 & J794e & 0.0625 \\
0.008334 & -0.375000 & 96.275253 & -60.58234 & 373.924 & J794e & 0.0625 \\
0.008334 & -0.625000 & 96.070511 & -60.81137 & 373.924 & J794e & 0.0625 \\
0.008334 & -0.875000 & 95.862831 & -61.04010 & 368.425 & J794e & 0.0625 \\
0.008335 & -1.125000 & 95.652138 & -61.26850 & 365.125 & J794e & 0.0625 \\
\enddata
\tablecomments{$RA$ is in decimal hours and $DEC$ is in degrees.
They are the central coordinates of small squares on the sky which satisfy
the selection criteria described in \S\ 3.
$l_{gal}$ and $b_{gal}$ are the Galactic longitude and latitude.
The coordinate system used is J2000. The exposure time (Expt) is in seconds.
Plate ID comes from DSS and area is expressed in sq. deg.}
\end{deluxetable}

\clearpage

\begin{deluxetable}{cccccc}
\tablecolumns{6}
\tablewidth{0pc}
\tablecaption{The Journal of the observations\label{jobs}}
\tablehead{
\colhead{Date} & \colhead{Telescope} & \colhead{Instrument} & \colhead{Slit} & \colhead{Resolution} & \colhead{Wavelength range}}
\startdata
October 1998 & 2.3m Bok & B\&C & $2^{``}.5$ & 20 \AA & 5000-9000 \AA \\
December 1999 & 2.3m Bok & B\&C & $2^{``}.5$ & 20 \AA & 5000-9000 \AA \\
March 2001 & 3.5m TNG & DOLORES & $1^{``}.5$ & 15 \AA & 4400-10000 \AA \\
March 2001 & 1.54m Danish & DFOSC & $1^{``}.5$ & 15 \AA & 3500-8500 \AA \\
September 2001 & 1.54m Danish & DFOSC & $1^{``}.5$ & 15 \AA & 3500-8500 \AA \\
\enddata
\tablecomments{2.3m Bok = Steward Observatory's 2.3m Bok Telescope at
Kitt Peak National Observatory (KPNO); 3.5m TNG = Italian 3.5m National
Telescope Galileo at Roque de Los Muchachos Observatory (ORM); 1.54m Danish =
Danish-ESO 1.54m Telescope at La Silla Observatory.
B\&C = Boller \& Chivens Spectrograph; DOLORES = Device Optimized for the
LOw RESolution; DFOSC = Danish Faint Object Spectrograph and Camera.}
\end{deluxetable}

\end{document}